\documentclass[a4paper,fleqn,usenatbib]{mnras}
\usepackage{newtxtext,newtxmath}
\usepackage[T1]{fontenc}
\usepackage{ae,aecompl}
\usepackage{graphicx}
\usepackage{pdflscape}

\newcommand{\kms}{\mbox{$\mathrm{km\,s^{-1}}$}}
\newcommand{\ms}{\mbox{$\mathrm{m\,s^{-1}}$}}
\newcommand{\MSUN}{\mbox{$\mathrm{M_{\odot}}$}}
\newcommand{\RSUN}{\mbox{$\mathrm{R_{\odot}}$}}

\title[White dwarf plus subgiants]{The White Dwarf Binary Pathways Survey  - IX. Three long period white dwarf plus subgiant binaries}

\author[S. G. Parsons et al.]{S.~G.~Parsons$^{1}$\thanks{s.g.parsons@sheffield.ac.uk},
M.~S.~Hernandez$^{2,3}$,
O.~Toloza$^{2,3}$,
M.~Zorotovic$^{4}$,
M.~R.~Schreiber$^{2,3}$,
\newauthor
B.~T.~G{\"a}nsicke$^{5}$,
F.~Lagos$^{5}$,
R.~Raddi$^{6}$,
A.~Rebassa-Mansergas$^{6,7}$,
J.~J.~Ren$^{8}$
\newauthor
and D.~Koester$^{9}$
\\
$^{1}$ Department of Physics and Astronomy, University of Sheffield,
Sheffield, S3 7RH, UK\\
$^{2}$ Departamento de F{\'i}sica, Universidad T{\'e}cnica Santa Mar{\'i}a, Avenida Espa{\~n}a 1680, Valpara{\'i}so, Chile\\
$^{3}$ Millennium Nucleus for Planet Formation, NPF, Valpara{\i}so, Av. Espa{\~n}a 1680, Chile \\
$^{4}$ Instituto de F{\'i}sica y Astronom{\'i}a de la Universidad de Valpara{\'i}so, Av. Gran Breta{\~n}a 1111, Valpara{\'i}so, Chile\\
$^{5}$ Department of Physics, University of Warwick, Coventry CV4 7AL, UK\\
$^{6}$ Departament de F{\'i}sica, Universitat Polit{\`e}cnica de Catalunya, c/Esteve Terrades 5, E-08860 Castelldefels, Spain\\
$^{7}$ Institute for Space Studies of Catalonia, c/Gran Capit{\`a} 2-4, Edif. Nexus 201, E-08034 Barcelona, Spain\\
$^{8}$ Key Laboratory of Space Astronomy and Technology, National Astronomical Observatories, Chinese Academy of Sciences, Beijing 100101, P. R. China\\
$^{9}$ Institut f{\"u}r Theoretische Physik und Astrophysik, University of Kiel, 24098 Kiel, Germany
}

\date{Accepted 2022 November 15. Received 2022 November 10; in original form 2022 July 15}

\pubyear{2022}

\begin{document}
\label{firstpage}
\pagerange{\pageref{firstpage}--\pageref{lastpage}}
\maketitle

\begin{abstract}

\noindent
Virtually all binaries consisting of a white dwarf with a non-degenerate companion can be classified as either close post-interaction systems (with orbital periods of a few days or less), or wide systems (with periods longer than decades), in which both components have effectively evolved as single stars. Binaries with periods between these two extremes can help constrain common envelope efficiency, or highlight alternative pathways towards the creation of compact binaries. To date such binaries have remained mostly elusive. Here we present three white dwarfs in binaries with evolved subgiant stars with orbital periods of 41, 52 and 461\,d. Using {\it Hubble Space Telescope} spectroscopy we find that all three systems contain low mass white dwarfs ($\leq$0.4\,{\MSUN}). One system, TYC\,8394$-$1331$-$1, is the inner binary of a hierarchical triple, where the white dwarf plus subgiant binary is orbited by a more distant companion star. These binaries were likely formed from a phase of stable but non-conservative mass transfer, as opposed to common envelope evolution. All three systems will undergo a common envelope phase in the future, but the two shorter period systems are expected to merge during this event, while the longest period system is likely to survive and create a close binary with two low mass white dwarfs. 

\end{abstract}


\section{Introduction}

Around one quarter of all solar-type stars are found in close binaries (P$_\mathrm{orb} < 10^4$\,days, \citealt{Moe19}), many of which are expected to interact with each other when the more massive member of the binary evolves off the main-sequence, often leading to a common envelope phase and a shrinking of the binary separation \citep{Paczynski76,Willems04}. This evolutionary pathway is thought to lead to the creation of compact binaries such as cataclysmic variables, double degenerate binaries and thermonuclear supernovae \citep{Webbink84}.

The common envelope phase itself, during which the core of the evolved star and its binary companion orbit within a shared envelope of material, is a brief but extremely complex process, which hydrodynamical models cannot yet fully recreate \citep{Ivanova13}. Instead, a simple energy equation with an efficiency, $\alpha_\mathrm{CE}$, is typically used in binary population models. High values of $\alpha_\mathrm{CE}$ imply efficient removal of the envelope and hence a small loss of orbital energy from the binary, resulting in relatively wide orbits after the common envelope. Conversely, low values of $\alpha_\mathrm{CE}$ imply inefficient envelope removal, causing a significant reduction in the binary separation and creating very close post-common envelope binaries.

Studies of close white dwarf plus M dwarf or brown dwarf binaries have shown that $\alpha_\mathrm{CE}$ is likely quite low for this class of systems, with values of $\alpha_\mathrm{CE}\simeq$~0.2-0.3 typically quoted \citep{Zorotovic10,Zorotovic14,Toonen13,Camacho14,Zorotovic22}. However, while such a low value of $\alpha_\mathrm{CE}$ is able to recreate this population of binaries, it is unclear how universal this is, particularly at higher stellar masses. Indeed, the common envelope phase in high-mass stars is often modelled as highly efficient, $\alpha_\mathrm{CE}\simeq1$ \citep[e.g.][]{Belczynski02}. Moreover, low efficiencies are unable to reproduce populations of double white dwarf binaries, since these systems need to remain at relatively wide separations after the first mass transfer phase in order to survive the second phase \citep{Nelemans20}. 

The challenge of reconstructing the evolution of double white dwarf binaries with the classical energy balance approach led \citet{Nelemans20} to suggest the so-called $\gamma$-formalism, which considers angular momentum balance (with some efficiency $\gamma$), as an alternative. This has the advantage that in some cases the separation of the two stars does not change much or can even increase as a result of mass transfer. However, this approach is not without its problems, as it was specifically designed to address the evolution of double white dwarf binaries it is unclear how applicable it is to other types of binaries undergoing common envelope evolution. Moreover, the $\gamma$-formalism appears to predict some systems that may violate energy conservation during their formation if only orbital and thermal energies are available \citep{Ivanova13}. As an alternative to this approach \citet{Webbink08} suggested that instead the first mass transfer event could be dynamically stable but non-conservative (as opposed to common envelope evolution, which is dynamically unstable and non-conservative), which can occur if the original mass ratio of the binary is close to one \citep{Ge20} and can lead to orbital expansion resulting in wider binaries \citep{Kawahara18,Lagos22}. Subsequent modelling has shown that the population of observed double white dwarf binaries can be recreated with a phase of stable and non-conservative mass transfer \citep{Woods12,Schreiber22} and also appears to be important for the creation of several other types of binary systems \citep[e.g.][]{Chen17,Vos19}. However, due to their large mass ratios, the progenitor systems of close white dwarf plus M dwarf binaries are not expected to experience stable and non-conservative mass transfer. White dwarfs with high mass companions are needed in order to investigate this process.

While there is much observational data for close white dwarf plus M dwarf binaries \citep[e.g.][]{Rebassa10,Rebassa16}, the situation is poorer for white dwarfs in close binaries with more massive A, F, G and K type stars (hereafter WD+AFGK binaries). In these binaries the white dwarfs are outshone by their companions at optical wavelengths by factors of hundreds or even thousands, only revealing themselves in the ultraviolet \citep{Parsons16}. WD+AFGK binaries are particularly important to study, since they represent the last common ancestor for a wide variety of compact binaries. Depending upon the post-mass transfer separation and stellar masses they may go on to form supersoft X-ray source systems and cataclysmic variables if they emerge at short periods \citep{Parsons15,Hernandez21,Hernandez22}, or symbiotic binaries and double white dwarf systems if they emerge at longer periods \citep{Zorotovic14}. Moreover, WD+AFGK binaries allow us to probe the common envelope efficiency at intermediate masses and investigate whether phases of stable but non-conservative mass transfer may have occurred in some of these systems and why this may be the case for some systems but not others.

The majority of WD+AFGK binaries studied to date have very short orbital periods, consistent with the same low value of $\alpha_\mathrm{CE}$ seen in white dwarf plus M dwarf binaries \citep{OBrien01,Parsons15,Krushinsky20,Hernandez21,Hernandez22}. However, a handful of longer period systems have also been identified, most of which were found as self-lensing binaries by the {\it Kepler} space mission \citep{Kruse14,Kawahara18,Masuda19}. The evolution of these longer period systems cannot be reconstructed using the same low common envelope efficiency \citep{Zorotovic14}. While it is possible to include other energy sources to help remove the envelope, such as recombination energy, it is not clear how substantial this would be and even this additional energy cannot help explain some of the widest systems.

In this paper we present 2MASS\,J18361702$-$5110583 (hereafter 2MASS\,J1836$-$5110), TYC\,6992$-$827$-$1 and TYC\,8394$-$1331$-$1, three new white dwarf binaries with orbital periods substantially longer than typical WD+AFGK binaries (P$_\mathrm{orb}>40$\,days) and evolved subgiant star companions (all three companions can be classified as K1$-$G9\,IV stars). Their long periods make them ideal laboratories to see if it is possible to create these systems via standard, low efficiency common envelope evolution or whether a stable but non-conservative mass transfer phase is required instead. With this in mind, we measure the stellar and binary parameters and reconstruct the past and future evolution of these systems.

\section{Observations and their reduction}

The three systems presented in this paper were first identified as candidate WD+AFGK binaries by \citet{Parsons16} who detected excess flux at ultraviolet wavelengths based on a combination of optical Radial Velocity Experiment (RAVE) data \citep{Kordopatis13} and ultraviolet Galaxy Evolution Explorer (GALEX) data \citep{Bianchi14}. Follow up ground-based spectroscopy revealed radial velocity variations in these systems, indicating binarity. Further ground-based spectroscopic observations were obtained in order to determine the orbital periods, while space-based ultraviolet data were taken in order to determine the properties of the white dwarfs. In this section we outline these ground- and space-based observations. 

\subsection{Optical Spectroscopy}

Ground-based data were collected with a range of telescopes and instruments over multiple years and comprise data taken in both visitor and service mode. Here we summarise the data collection and reduction procedure for each instrument.

\subsubsection{Du Pont echelle}

We used the high resolution echelle spectrograph (1 arcsec slit, R$\simeq$40\,000, covering the wavelength range 3700{\AA} to 7000{\AA}) on the 2.5-m Du Pont telescope located at Las Campanas Observatory, Chile  to obtain two spectra of 2MASS\,J1836$-$5110 (on 2014 June 1 and 2) and four spectra of TYC\,6992$-$827$-$1 (on 2014 June 2, 2015 January 3, 4 and 5). Each science observation was bracketed by ThAr spectra to correct for instrumental drift. Standard image reductions were performed and the spectra optimally extracted and wavelength calibrated using the Collection of Elemental Routines for Echelle Spectra (CERES) package \citep{brahm17}. 

\subsubsection{FEROS}

High resolution spectra for all three systems were obtained with the FEROS echelle spectrograph (R$\simeq$48\,000) on the 2.2-m Telescope at La Silla, Chile \citep{kaufer98}. FEROS covers the wavelength range from $\simeq$3500{\AA} to $\simeq$9200\AA. Observations were performed in Object-Calibration mode where one fibre is placed on the target while the other feeds light from a ThAr+Ne calibration lamp permitting velocity measurements to extremely high precision ($\simeq$10{\ms}) and allowed us to correct for instrumental drift throughout the night. FEROS data were obtained over multiple nights covering several years and all were reduced using the CERES package.

\subsubsection{CHIRON}

Both TYC\,6992$-$827$-$1 and TYC\,8394$-$1331$-$1 were observed with the high resolution echelle spectrometer CHIRON \citep{tokovinin13} on the 1.5-m SMARTS telescope at Cerro Tololo, Chile. We used $3\times1$ binning resulting in R$\simeq$40\,000 covering a wavelength range of 4150{\AA} to 8800{\AA}. CHIRON observations were performed in service mode over many nights in both 2014 and 2015. These data are automatically reduced by the CHIRON team using standard reduction methods. 

\subsubsection{UVES}

All three targets were observed with UVES \citep{dekker00}, a high resolution echelle spectrograph mounted on the 8.2-m European Southern Observatory Very Large Telescope at Cerro Paranal, Chile, as part of a poor weather program. We used the dichroic 1 setup (390+564) with a 0.7 arcsec slit, resulting in R$\simeq$50\,000 and covering the wavelength range 3260{\AA} to 6680{\AA} with a small gap between 4540{\AA} and 4580{\AA}. Typical exposure times of 300\,s were used and the data were reduced using the UVES data reduction pipeline (version 5.8.2) using standard reduction methods within {\sc esoreflex}.

\subsubsection{X-Shooter}

A single observation of TYC\,8394$-$1331$-$1 was obtained with the medium resolution spectrograph X-Shooter \citep{Vernet11}, mounted on the 8.2-m European Southern Observatory Very Large Telescope on the night of 2021 June 6. X-Shooter covers both the optical and near-infrared wavelength ranges using three arms, the UVB (3000$-$5600\,{\AA}), VIS (5600$-$10,000\,{\AA}) and NIR (10\,000$-$24\,000\,{\AA}). Separate slit widths can be set for each arm and our observations were performed with slit widths of 1.0, 0.9 and 0.9 arcsec in the UVB, VIS and NIR arms respectively, giving R$\simeq$5\,000. Observations were performed in STARE mode, with exposure times of 120\,s in the UVB and VIS arms and 60\,s in the NIR arm. The data were reduced using the X-shooter reduction pipeline (version 3.5.0) within {\sc esoreflex}.

\subsection{Ultraviolet Spectroscopy}

We obtained far-ultraviolet (FUV) spectroscopy of all three systems in order to confirm the presence of a white dwarf companion and measure the stellar parameters. We used the Space Telescope Imaging Spectrograph (STIS, \citealt{Kimble98}) on-board the {\it Hubble Space Telescope} ({\it HST}) as part of program GO 16224. TYC\,8394$-$1331$-$1 was observed on 2021 May 27 over one spacecraft orbit, with a total exposure time of 2325\,s. TYC\,6992$-$827$-$1 was observed on 2022 January 10 for two orbits, for a total exposure time of 4980\,s. 2MASS\,J1836$-$5110 was observed on 2021 April 9 for two orbits, for a total exposure time of 5244\,s. Data were acquired with the MAMA detector and the G140L grating. The spectra cover a wavelength range of 1150$-$1730{\AA} with a resolving power between 960$-$1440. The spectra were reduced and wavelength calibrated following the standard STIS pipeline \citep{Sohn19}.

\section{Fitting procedures}

\subsection{Radial velocity measurements}

Radial velocities were computed from all our echelle spectra using cross-correlation against a binary mask representative of a G2-type star, which is the closest spectral type to our targets (K$0\pm1$, i.e. K1$-$G9) available (see \citealt{brahm17} for more information). While this technique is specifically designed to work with main-sequence stars, we have found that the results are also reliable for subgiant stars, provided they are relatively slow rotators (which is the case for all three systems presented in this paper). In general this technique yields velocities to a precision of $\sim$0.1 {\kms} or better. However, due to the unstable nature of the Du Pont and CHIRON spectrographs, as well as the uncertain systematic errors introduced in our UVES and X-shooter reductions, we placed lower limits of 0.5\,{\kms} on the precision of our velocity measurements from these instruments. This limit is based on previous experience with these instruments. The full list of radial velocity measurements for each system are given in Tables~\ref{tab:2mass1836_vels}, \ref{tab:tyc6992_vels} and \ref{tab:tyc8394_vels} in the appendix.

\subsection{Orbital period measurements}

\begin{figure*}
  \begin{center}
    \includegraphics[width=\textwidth]{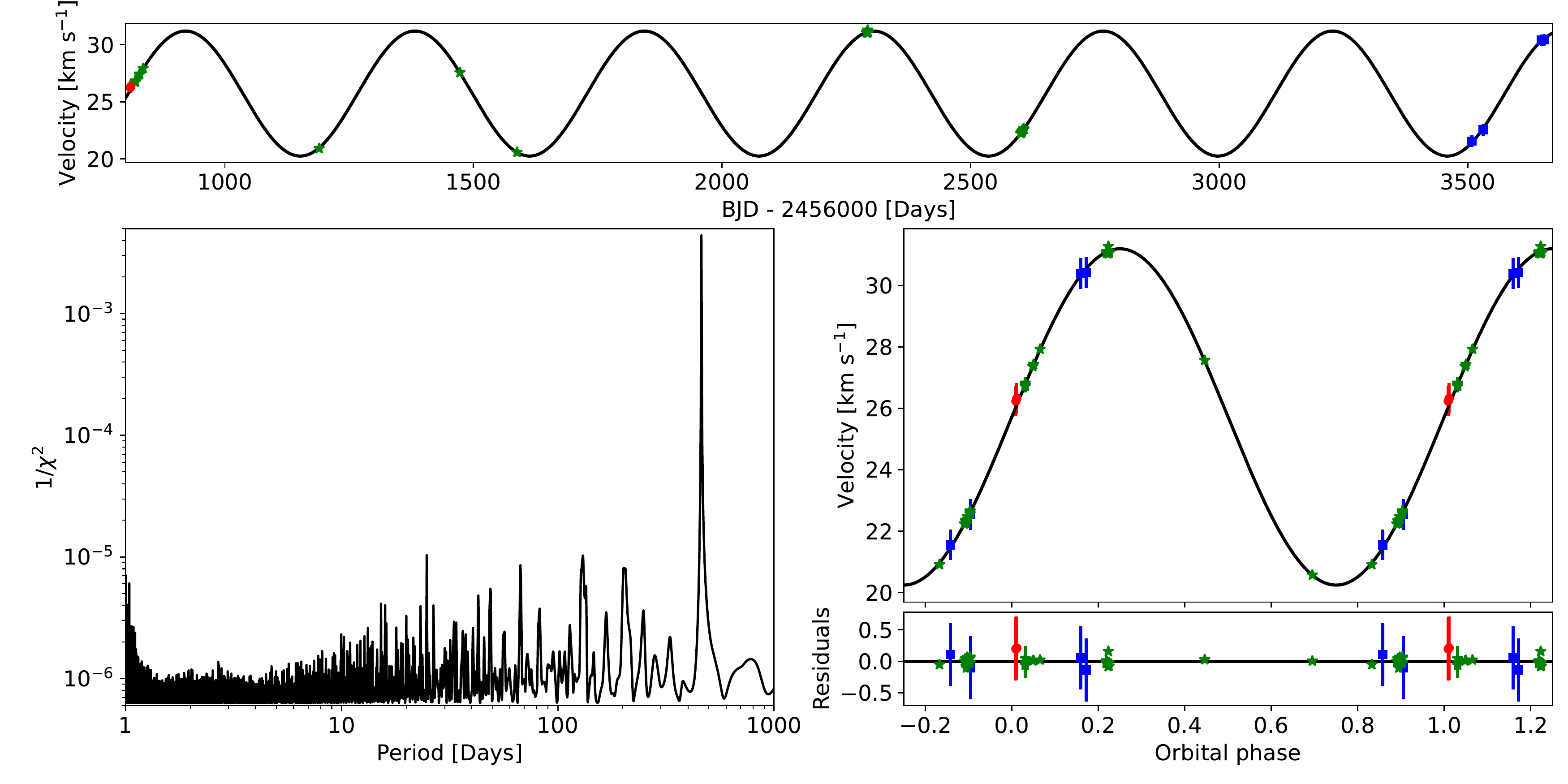}
    \caption{Radial velocity fit for 2MASS\,J1836$-$5110. The top panel shows the full data set and Keplerian orbit fit (Du Pont echelle -- red dots, FEROS -- green stars, UVES -- blue squares). The periodogram is shown in the bottom-left and the phase-folded data are shown in the bottom-right panel (phase zero corresponds to the time of inferior conjunction of the subgiant star).}
  \label{fig:2mass1836_rvs}
  \end{center}
\end{figure*}

\begin{figure*}
  \begin{center}
    \includegraphics[width=\textwidth]{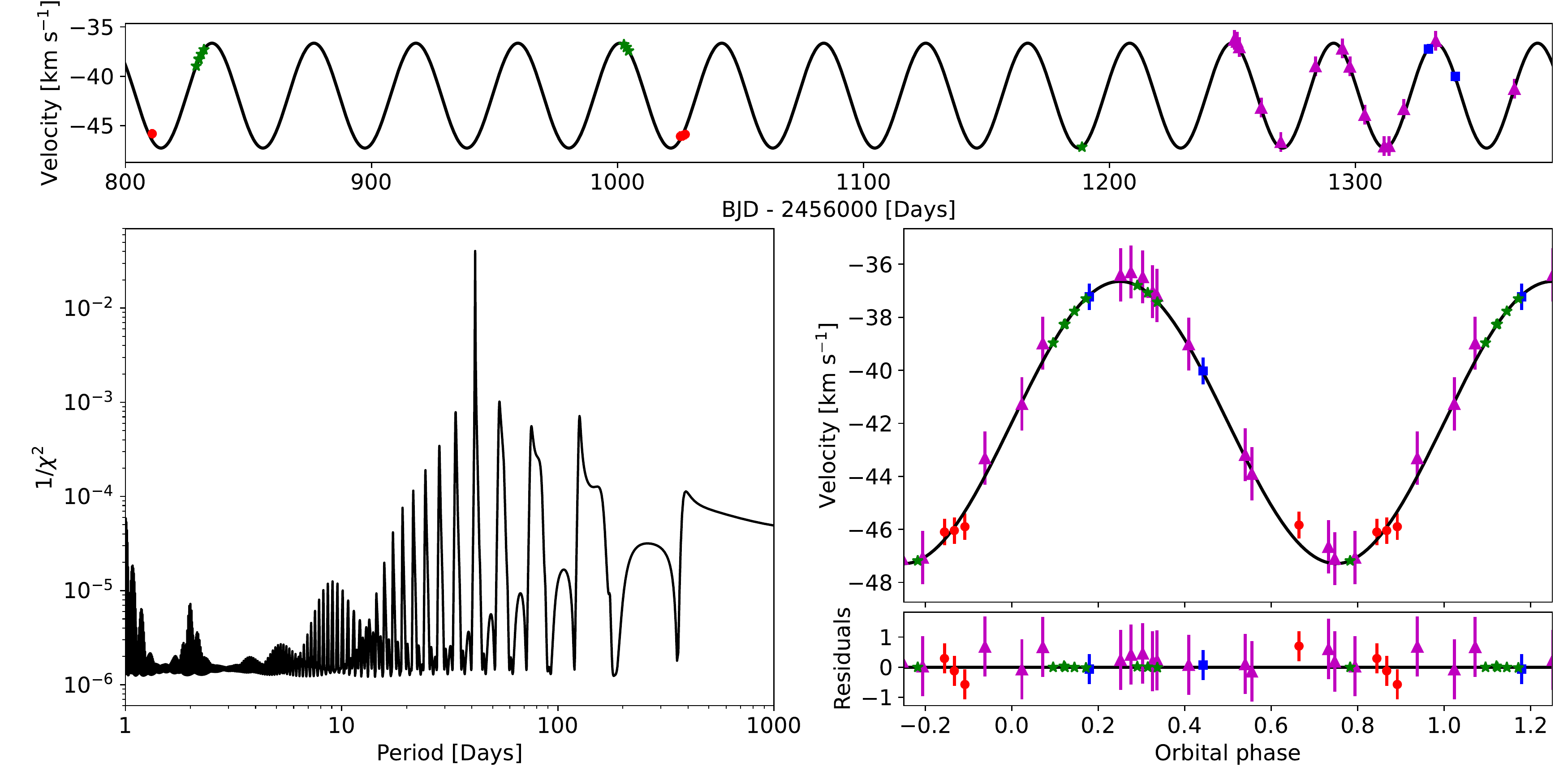}
    \caption{Same as Figure~\ref{fig:2mass1836_rvs} but for TYC\,6992$-$827$-$1 (Du Pont echelle -- red dots, FEROS -- green stars, UVES -- blue squares, CHIRON -- magenta triangles).}
  \label{fig:tyc6992_rvs}
  \end{center}
\end{figure*}

\begin{figure*}
  \begin{center}
    \includegraphics[width=\textwidth]{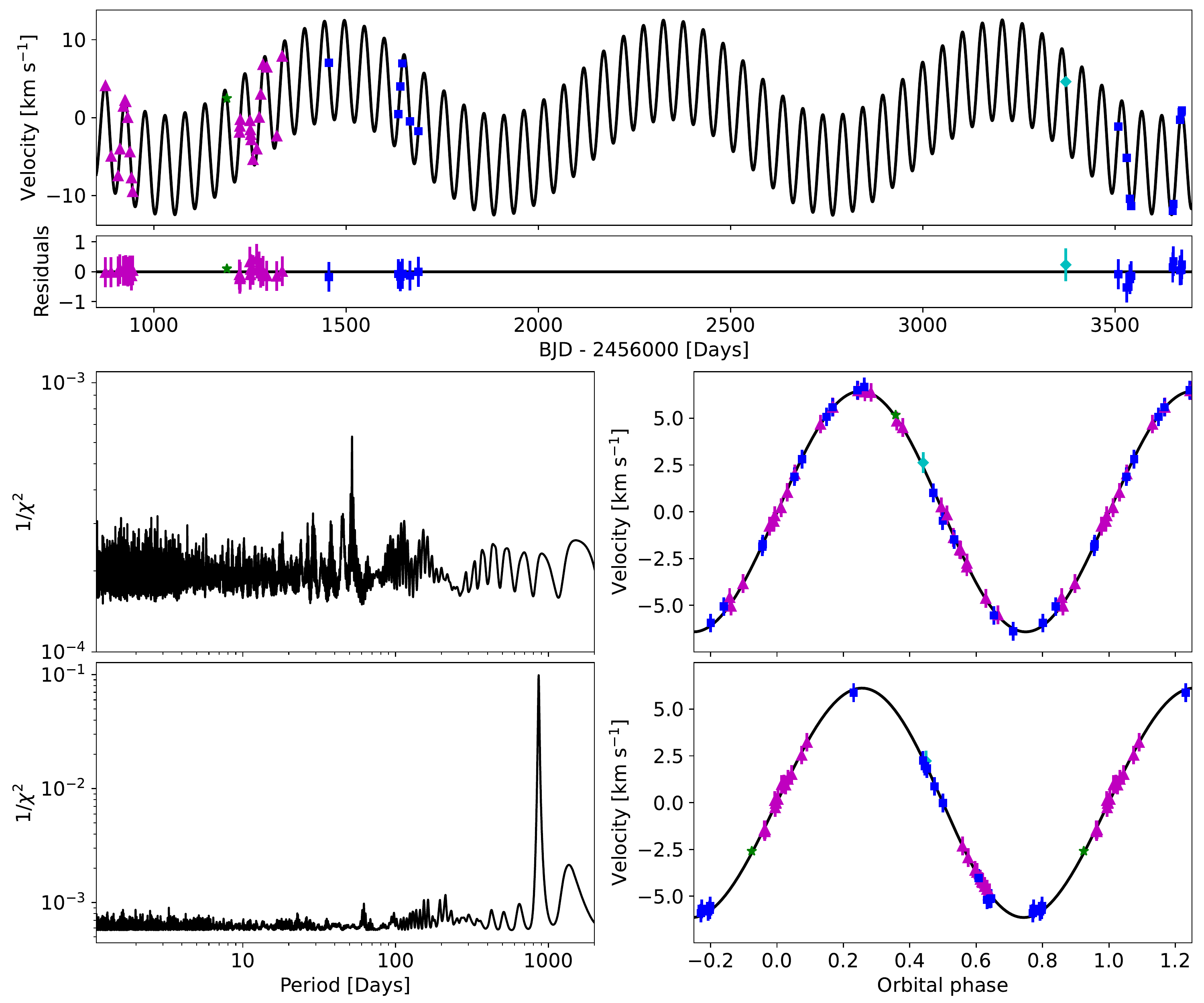}
    \caption{Radial velocity fit for TYC\,8394$-$1331$-$1 using a combination of two Keplerian orbits. The top panel shows the full data set and fit ((FEROS -- green stars, UVES -- blue squares, CHIRON -- magenta triangles, X-shooter -- cyan diamonds). The centre-left panel shows the periodogram of the full dataset. The centre-right panel shows the data folded on the shorter period signal, with the longer period signal removed. The lower-left panel shows the periodogram with the strongest signal from the original periodogram subtracted off. The bottom-right panel shows the data folded on the longer period signal, with the shorter period signal removed.}
  \label{fig:tyc8394_rvs}
  \end{center}
\end{figure*}

\begin{table}
 \centering
  \caption{Best fit Keplerian orbit parameters and uncertainties for 2MASS\,J1836$-$5110 and TYC\,6992$-$827$-$1. Parameters directly constrained from the data are indicated as "fitted", while parameters derived from these are listed as "derived".}
  \label{tab:rvfit1}
  \begin{tabular}{@{}lcc@{}}
    \hline
    Parameter & 2MASS\,J1836$-$5110 & TYC\,6992$-$827$-$1 \\
    \hline
    \multicolumn{3}{c}{Fitted:} \\
    P$_\mathrm{orb}$ [days] & $461.48\pm0.04$ & $41.45\pm0.01$ \\
    T$_\mathrm{conj}$ [BJD] & $2458189.9 \pm 0.1$ & $2457073.50\pm0.06$ \\
    $\sqrt{e}\cos{\omega}$  & $-0.075\pm0.005$ & $-0.06\pm0.06$ \\
    $\sqrt{e}\sin{\omega}$  & $-0.151\pm0.005$ & $0.08\pm0.06$ \\
    $\ln{K}$ [$\kms$]       & $1.700\pm0.001$ & $1.683\pm0.006$ \\
    \multicolumn{3}{c}{Derived:} \\
    $\gamma$ [$\kms$]       & $25.717\pm0.004$ & $-41.969\pm0.005$ \\
    $e$                     & $0.028\pm0.001$  & $0.013\pm0.06$ \\
    $K$ [$\kms$]            & $5.476\pm0.005$  & $5.38\pm0.03$ \\
    \hline
  \end{tabular}
\end{table}

\begin{table}
 \centering
  \caption{Best fit parameters and uncertainties for the double Keplerian orbit of TYC\,8394$-$1331$-$1. The subscripts 1 and 2 refer to the inner and outer orbits respectively.}
  \label{tab:rvfit2}
  \begin{tabular}{@{}lc@{}}
    \hline
    Parameter & value \\
    \hline
    \multicolumn{2}{c}{Fitted:} \\
    P$_\mathrm{orb,1}$ [days]  & $51.851\pm0.009$ \\
    T$_\mathrm{conj,1}$ [BJD]  & $2458337.8\pm0.3$ \\
    $\sqrt{e_1}\cos{\omega_1}$ & $0.04\pm0.10$ \\
    $\sqrt{e_1}\sin{\omega_1}$ & $0.07\pm0.08$ \\
    $\ln{K_1}$ [$\kms$]        & $1.86\pm0.02$ \\
    P$_\mathrm{orb,2}$ [days]  & $863\pm3$ \\
    T$_\mathrm{conj,2}$ [BJD]  & $2458549\pm6$ \\
    $\sqrt{e_2}\cos{\omega_2}$ & $-0.03\pm0.10$ \\
    $\sqrt{e_2}\sin{\omega_2}$ & $0.1\pm0.1$ \\
    $\ln{K_2}$ [$\kms$]        & $1.80\pm0.03$ \\
    \multicolumn{2}{c}{Derived:} \\
    $\gamma$ [$\kms$]         & $35.4\pm0.1$ \\
    $e_1$                     & $0.02\pm0.02$ \\
    $K_1$ [$\kms$]            & $6.4\pm0.1$ \\
    $e_2$                     & $0.03\pm0.02$ \\
    $K_2$ [$\kms$]            & $6.1\pm0.2$ \\
    \hline
  \end{tabular}
\end{table}

Initial estimates of the orbital periods were made by fitting a constant plus sine wave to the velocity measurements over a range of periods and computing the $\chi^2$ of the resulting fit at each period (i.e. a periodogram). This approach works well for orbits that are very close to circular. These initial estimates were then used as starting points for fitting the Keplerian orbits to the measured radial velocities using the Python package {\sc radvel} \citep{Fulton18}. 

We allowed the eccentricity to vary (where $\sqrt{e}\cos{\omega}$ and $\sqrt{e}\sin{\omega}$ are the fitted parameters, $e$ is the eccentricity and $\omega$ is the argument of periapsis) and fitted the radial velocity as $\ln{K}$, where $K$ is the semi-amplitude (i.e. the velocity varies over time $t$ as $K\sin{t}$), which is particularly useful when $K$ is large compared to the uncertainty on each measurement. Along with these parameters we also fitted P$_\mathrm{orb}$, the orbital period and T$_\mathrm{conj}$, the time of inferior conjunction of the subgiant star. The systemic velocity, $\gamma$, was solved analytically during the fit and so was not a free parameter\footnote{\url{http://cadence.caltech.edu/~bfulton/share/Marginalizing_the_likelihood.pdf}}.

The distributions of our model parameters were found using the Markov chain Monte Carlo (MCMC) method as implemented within {\sc radvel}. We placed a uniform prior on the eccentricity to ensure it stayed between zero and one and a uniform prior on T$_\mathrm{conj}$ to prevent it from changing by more than one orbital cycle.

For TYC\,8394$-$1331$-$1 it was immediately clear that a single Keplerian orbit was insufficient to model the radial velocity data. Therefore, this system was modelled with a combination of two orbits. An initial estimate for the period of the outer orbit was made by subtracting the best fit sinusoid from the radial velocity data and fitting the residuals with an additional sine wave over a range of periods. The best fit value was then used as the starting period for the outer orbit for the {\sc radvel} fit. For this system both orbits were then fitted simultaneously using the same parameters for each orbit detailed above.

There were no clear additional orbits in either 2MASS\,J1836$-$5110 or TYC\,6992$-$827$-$1. The best fit parameters for these two systems are listed in Table~\ref{tab:rvfit1}, while the best fit orbital parameters for TYC\,8394$-$1331$-$1 are given in Table~\ref{tab:rvfit2}. The periodograms, best fits and residuals are shown in Figures~\ref{fig:2mass1836_rvs}, \ref{fig:tyc6992_rvs} and \ref{fig:tyc8394_rvs} for 2MASS\,J1836$-$5110, TYC\,6992$-$827$-$1 and TYC\,8394$-$1331$-$1 respectively.

\subsection{Subgiant parameters}

\begin{figure*}
  \begin{center}
    \includegraphics[width=\textwidth]{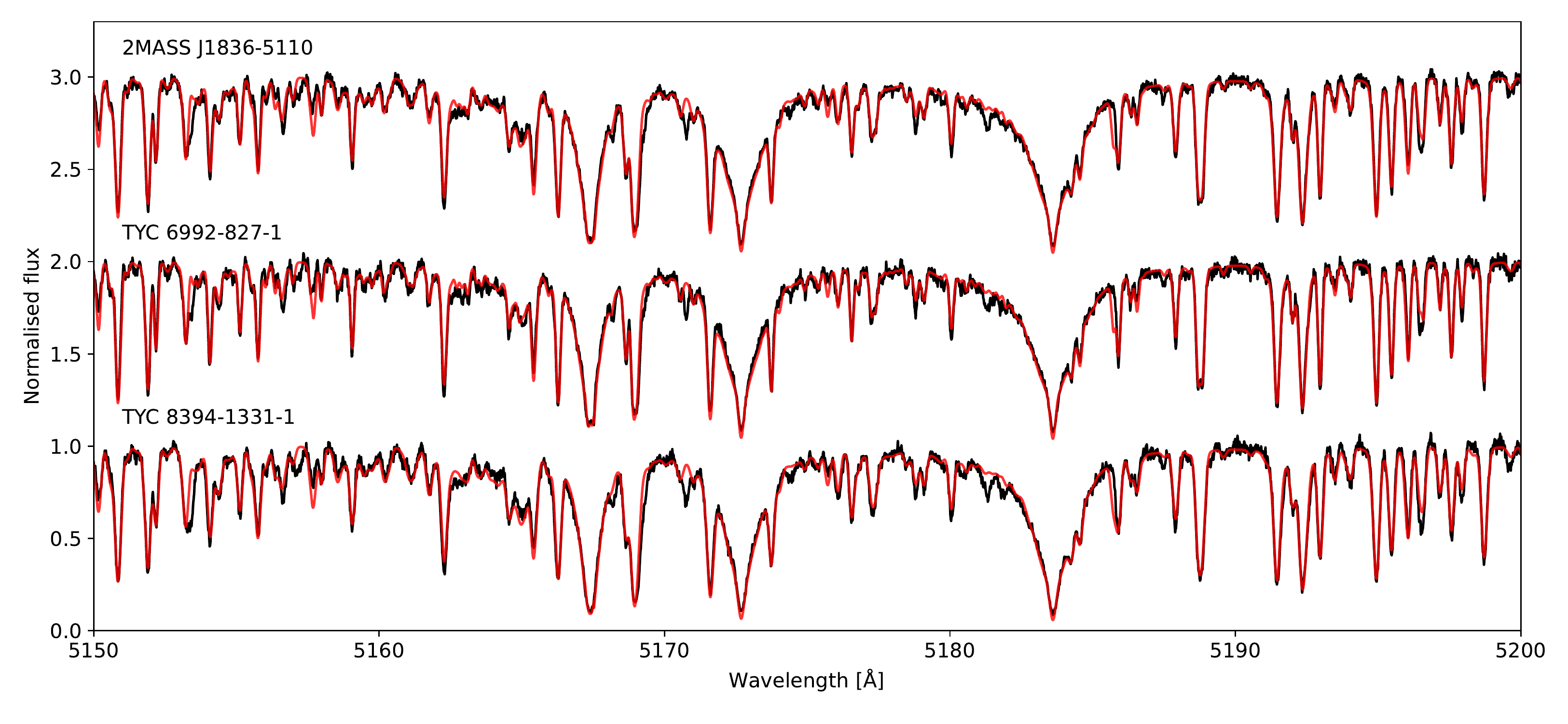}
    \caption{Example spectra (black) and model atmosphere fits (red) for all three systems. In each case a single UVES spectrum is shown. The fits cover the entire spectral range, but we plot just the region around the Mg\,{\sc i} triplet for clarity.}
  \label{fig:specs}
  \end{center}
\end{figure*}

All three objects presented in this paper were originally selected as ultraviolet excess main-sequence stars ($\log{g}>3.5$) from RAVE DR4 \citep{Kordopatis13}, before any {\it Gaia} data releases \citep{Parsons16}. However, the {\it Gaia} DR2 parallaxes revealed that these objects are slightly evolved and subsequent RAVE data releases (which factor in {\it Gaia} parallaxes) have revised down the surface gravities of all these objects \citep{Steinmetz20}. Our extensive follow-up spectroscopy and the release of {\it Gaia} DR3 allowed us to place more precise constraints on these stars than the most recent RAVE DR6 values and in this section we detail how this was achieved.

We determined the stellar parameters of the subgiant stars in all three binaries in two steps. Firstly, we fitted the high resolution optical spectra of each object to measure the effective temperature (T$_\mathrm{eff,SG}$, where SG refers to subgiant), surface gravity ($\log{g}_\mathrm{SG}$), metallicity ([M/H]) and rotational broadening ($v_\mathrm{SG}\sin{i}$) using the spectral analysis software {\sc ispec} \citep{Blanco14,Blanco19}. We decided to fit all of the UVES spectra for each object, since these data are generally the highest signal-to-noise ratio spectra and they cover a wide wavelength range.

We used the {\sc moog}\footnote{\url{http://www.as.utexas.edu/~chris/moog.html}} radiative transfer code and the MARCS GES/APOGEE model atmosphere grid \citep{Gustafsson08}. {\sc ispec} uses a least-squares algorithm to minimize the difference between the synthetic and observed spectra. In each iteration, the algorithm varies one free parameter at a time in order to determine in which direction it should move \citep{Blanco14}. In order to speed up the fitting, {\sc ispec} interpolates between the pre-computed MARCS GES/APOGEE models if the values lie inside the grid, otherwise a spectrum is synthesised using {\sc moog}. Micro- and macro-turbulences were not fitted but were determined using empirical relations \citep{Blanco14b}. The linear limb darkening coefficient was fixed at 0.6 and the resolution was also fixed at 50,000. All spectra were first corrected for radial velocity shifts based on the measured values and the velocity was then fixed at zero during the spectral fitting. Initial values for T$_\mathrm{eff,SG}$, $\log{g}_\mathrm{SG}$, [M/H] were set using the RAVE DR6 parameters \citep{Steinmetz20}, while $v_\mathrm{SG}\sin{i}$ was initiated at 5{\kms}. Additional fits with the initial parameters altered by T$_\mathrm{eff,SG}\pm100$\,K, $\log{g}_\mathrm{SG}\pm0.25$\,dex, [M/H]$\pm0.1$ and $v_\mathrm{SG}\sin{i}\pm5$\,{\kms} were also performed, to ensure that the fits always converged to the same best fit values. Examples of the best fitting spectral models are shown in Figure~\ref{fig:specs}, in a small range around the Mg\,{\sc i} triplet (although the entire UVES spectral range was included in the fit, with the exception of any wavelength ranges affected by telluric lines).

The best fit parameters for a specific star varied from spectrum to spectrum by an amount comparable to the uncertainties on the individual fits. We therefore combined the results from all UVES spectral fits and adopted the inverse-variance weighted mean and variance as the final best fit parameters and their uncertainties. These are listed in Table~\ref{tab:sgparams} for all three systems.

The second step in determining the subgiant parameters was to fit their spectral energy distributions (SEDs) in order to measure their radii (R$_\mathrm{SG}$) and, when combined with $\log{g}_\mathrm{SG}$, hence their masses (M$_\mathrm{SG}$). We obtained {\it Gaia} DR3 $G_\mathrm{BP}$, $G$ and $G_\mathrm{RP}$ data \citep{Gaia21}, 2MASS $J$, $H$ and $K_\mathrm{S}$ band data \citep{Skrutskie06} and WISE $W1$ and $W2$ photometry \citep{Cutri12} for all three systems. We fitted these photometric data with solar metallicity BT-Settl models \citep{Allard11}. 

\begin{table}
 \centering
  \caption{Best fit stellar parameters for the subgiant stars in our binaries}
  \label{tab:sgparams}
  \tabcolsep=0.05cm
  \begin{tabular}{@{}lccc@{}}
    \hline
    Parameter & \hspace{-7mm} 2MASS\,J1836$-$5110 & TYC\,6992$-$827$-$1 & TYC\,8394$-$1331$-$1 \\
    \hline
    \multicolumn{4}{c}{From spectral fit:} \\
    T$_\mathrm{eff,SG}$ [K]       & $5050\pm50$    & $5250\pm50$    & $5150\pm20$ \\
    $\log{g}_\mathrm{SG}$ [dex]   & $3.48\pm0.05$  & $3.48\pm0.04$  & $3.06\pm0.02$ \\
    $[\mathrm{M/H}]$ [dex]        & $-0.05\pm0.07$ & $-0.10\pm0.10$ & $0.03\pm0.15$ \\
    $v_\mathrm{SG}\sin{i}$ [\kms] & $4.7\pm0.5$    & $2.6\pm0.2$    & $6.2\pm0.6$ \\
    \multicolumn{4}{c}{From {\it Gaia} DR3:} \\
    $\varpi$ [mas]                & $1.09\pm0.02$  & $1.98\pm0.06$  & $1.44\pm0.07$ \\
    \multicolumn{4}{c}{From STILISM or \citet{Schlafly11}$^*$:} \\
    $E(B-V)$ [mag]                & $0.06\pm0.02$  & $<0.02^*$  & $0.04\pm0.02$ \\
    \multicolumn{4}{c}{From SED fit:} \\
    R$_\mathrm{SG}$ [\RSUN]       & $3.54\pm0.07$  & $3.45\pm0.12$  & $5.57\pm0.24$ \\
    \multicolumn{4}{c}{Derived:} \\
    M$_\mathrm{SG}$ [\MSUN]       & $1.38\pm0.16$  & $1.31\pm0.14$  & $1.31\pm0.12$ \\
    \hline
  \end{tabular}
\end{table}

Models were generated for a given $\log{g}_\mathrm{SG}$ and T$_\mathrm{eff,SG}$, the model fluxes (which are $4\pi\times F_\mathrm{Eddington}$) were then scaled by a factor of (R$_\mathrm{SG}$/D)$^2$, where D is the distance, and were reddened by a factor $E(B-V)$. The model flux in each photometric band was then calculated and compared to the observed values. The fit was performed using the MCMC method as implemented in the {\sc emcee} python package \citep{Foreman13}. We placed Gaussian priors on $\log{g}_\mathrm{SG}$ and T$_\mathrm{eff,SG}$ based on the fit to the high resolution spectra. A prior was also placed on the reddening, based on the value for each system from the STILISM reddening map \citep{Capitanio17}, the reddening was also forced to be a positive value. For TYC\,6992$-$827$-$1 the maps do not extend far enough to cover this object, so we place an upper limit on the reddening of this object based on the maximum reddening set by \citet{Schlafly11}. Finally, we used the {\it Gaia} DR3 parallaxes to calculate the distances (via parallax inversion, note that all three systems have $\varpi/\sigma_\varpi>20$) and placed a Gaussian prior on the parallax based on the {\it Gaia} DR3 value.

\begin{figure}
  \begin{center}
    \includegraphics[width=\columnwidth]{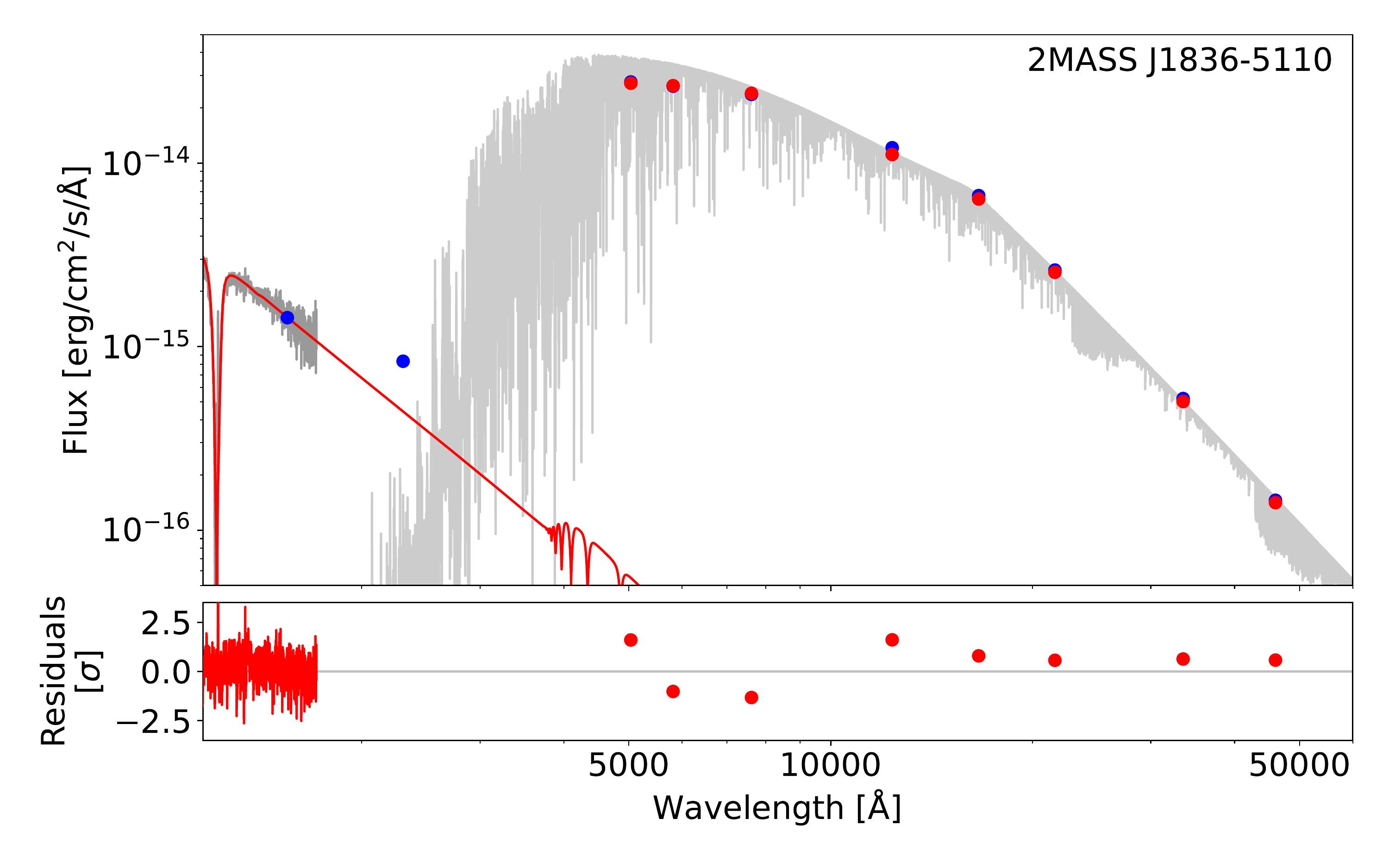}
    \includegraphics[width=\columnwidth]{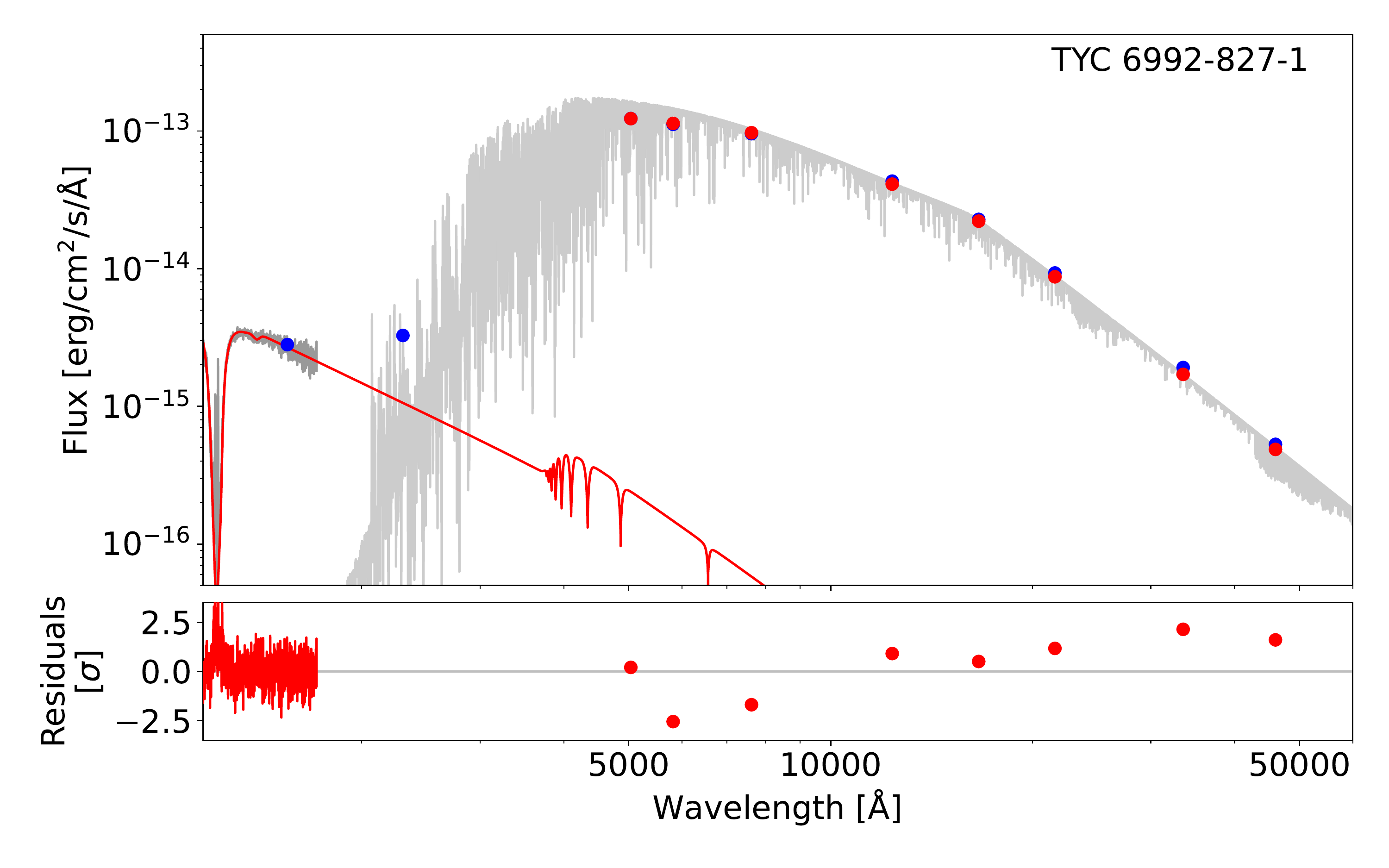}
    \includegraphics[width=\columnwidth]{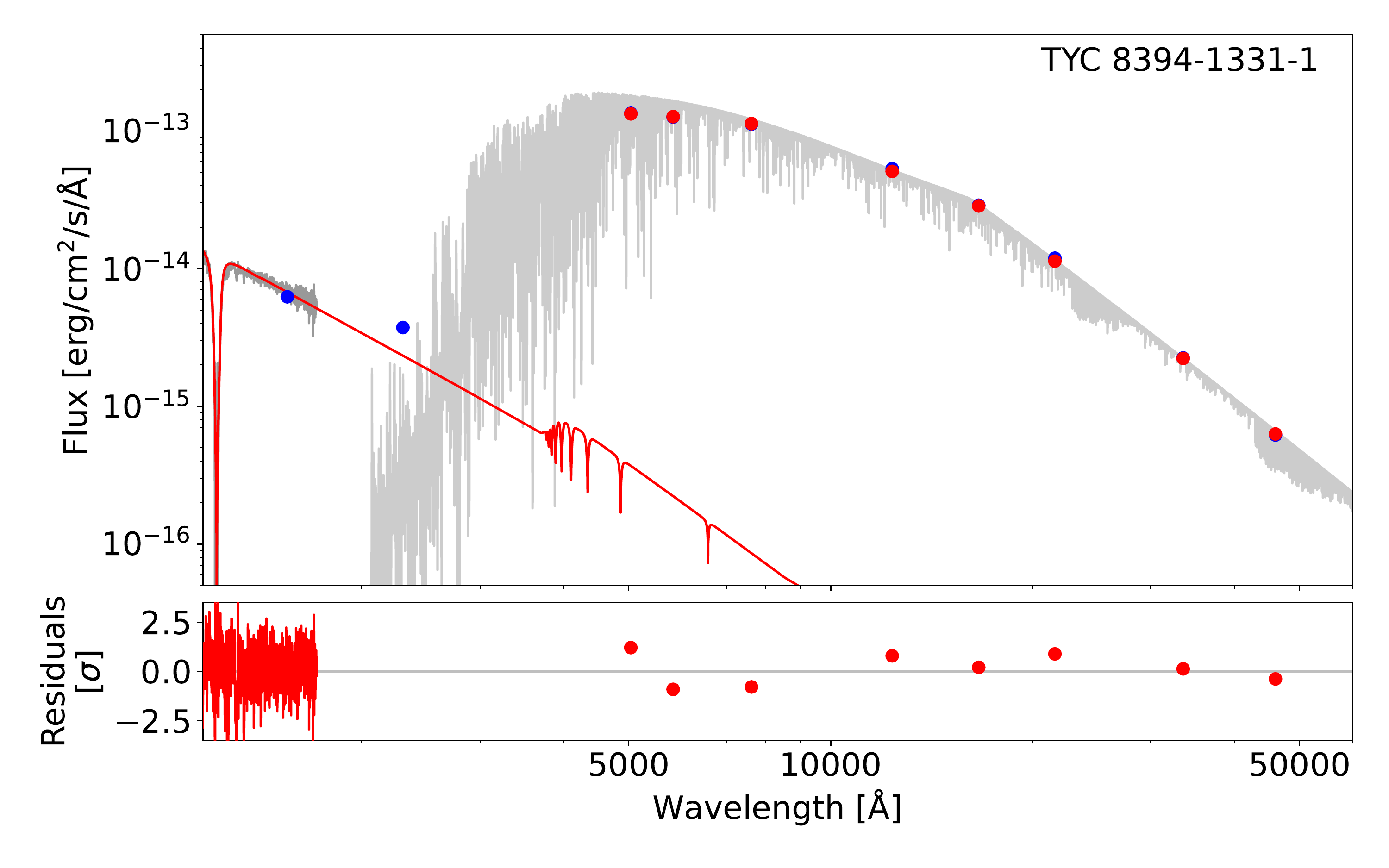}
    \caption{SED fits to the subgiant stars in the three systems presented in this paper. The blue points show the GALEX NUV and FUV, {\it Gaia} $G_\mathrm{BP}$, $G$ and $G_\mathrm{RP}$, 2MASS $J$, $H$ and $K_\mathrm{S}$ and WISE $W1$ and $W2$ band measurements (error bars are plotted but are too small to be visible). The best fit model spectra to the subgiant stars are shown in light grey and the model fluxes in the observed bands are shown as red points (the GALEX data were not included in the fit). We also show the {\it HST}/STIS spectra of the white dwarfs in dark grey and the best fit white dwarf spectrum in red. The lower panels show the residuals to both the white dwarf and subgiant star fits.}
  \label{fig:seds}
  \end{center}
\end{figure}

The best fit SEDs are shown in Figure~\ref{fig:seds} and the best fit reddening, radii and masses (from $\log{g}_\mathrm{SG}$) are listed in Table~\ref{tab:sgparams}. The MCMC parameter distributions are also shown in Figure~\ref{fig:sed_corner} in the Appendix. Based on the fitted temperatures, all three objects have the spectral classification of K$0\pm1$\,IV. However, we add a note of caution here, since we will see in the next section that the white dwarf cooling ages in 2MASS\,J1836$-$5110 and TYC\,6992$-$827$-$1 are comparable to the thermal timescale of the subgiant stars. Therefore, these stars may not be fully relaxed to their equilibrium radii. Since the theoretical models are based on stars in equilibrium, it is possible that additional systematic errors exist on the subgiant parameters in these two systems. Unfortunately, since we do not know how far from equilibrium these stars are and theoretical models do not exist for such objects, this is the best we can do at present.

\begin{figure}
  \begin{center}
    \includegraphics[width=\columnwidth]{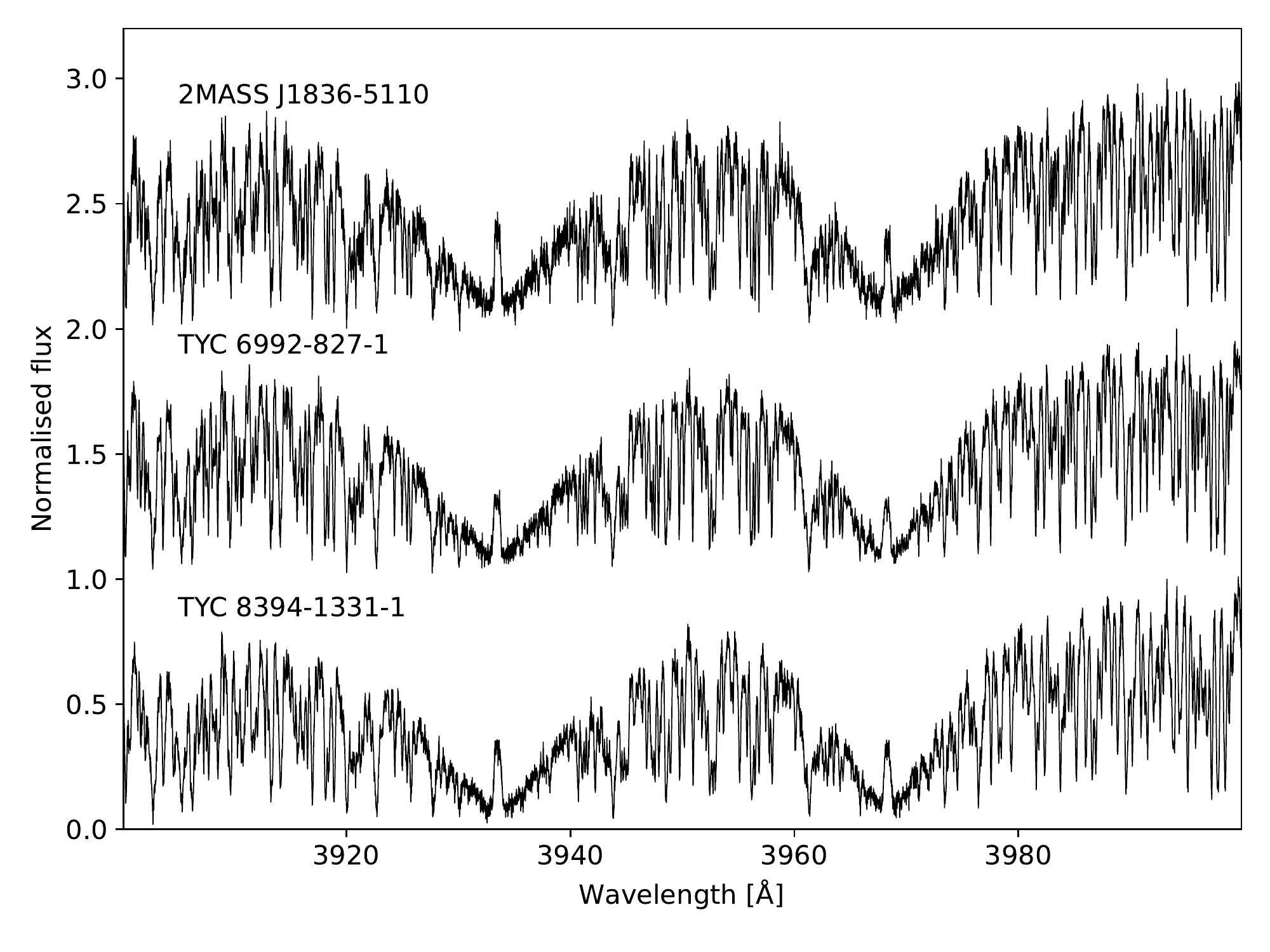}
    \caption{UVES spectra around the Ca\,{\sc ii} H and K lines, showing clear emission lines in all three systems (which track the motion of the subgiant stars), likely due to chromospheric activity.}
  \label{fig:activity}
  \end{center}
\end{figure}

The GALEX FUV and NUV measurements are also indicated in Figure~\ref{fig:seds}, although these were not included in the SED fits (since the white dwarf contributes a non-negligible amount of flux at these wavelengths). In all three cases the NUV measurements show a clear excess, even after accounting for the white dwarf contribution. A similar effect has been observed in other WD+AFGK binaries \citep{Hernandez22,Hernandez22b} and was attributed to chromospheric emission from the rapidly rotating and highly active main-sequence star components. While there is no evidence of rapid rotation in the subgiant stars presented in this paper, there is clear evidence of chromospheric activity in the form of Ca\,{\sc ii} H and K emission from the subgiants in all three systems (see Figure~\ref{fig:activity}). Chromospheric emission is known to cause excess flux at NUV wavelengths. For example NUV excesses have been detected in rapidly rotating red giants, \citealt{Dixon20}, and even the subgiant star in the wide white dwarf K subgiant binary UCAC2\,46706450 \citep{Werner20}, and since chromospheric emission is not included in the BT-Settl models used to model the SEDs of our subgiant stars this is likely the cause of our under predicted NUV fluxes.

Chromospheric emission may also be the cause of the slight overestimate of the {\it Gaia} $G_\mathrm{BP}$ fluxes seen in Figure~\ref{fig:seds}. However, excluding these measurements from the SED fit does not significantly change the subgiant parameters.

\subsection{White dwarf parameters}

\begin{figure}
  \begin{center}
    \includegraphics[width=\columnwidth]{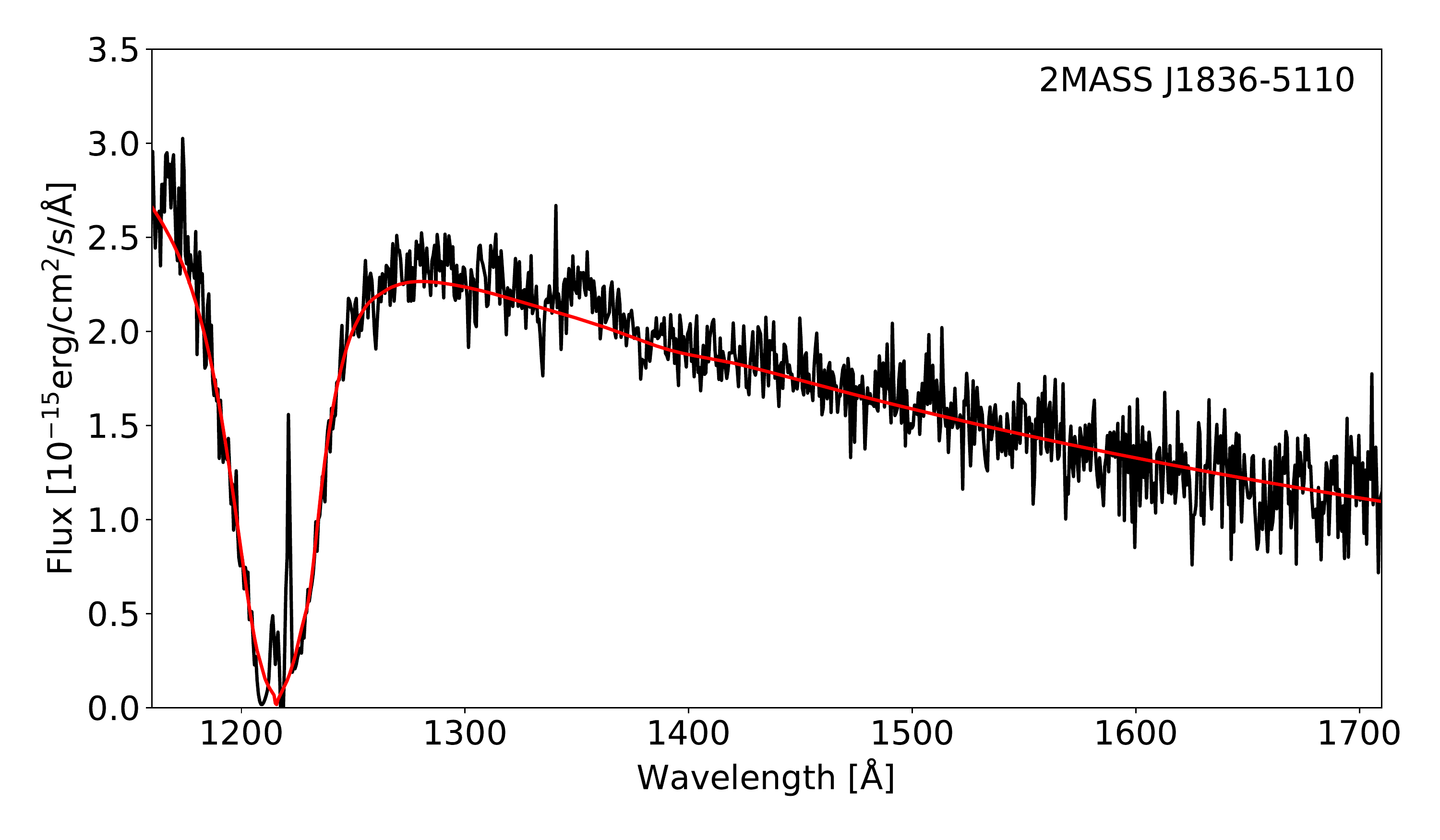}
    \includegraphics[width=\columnwidth]{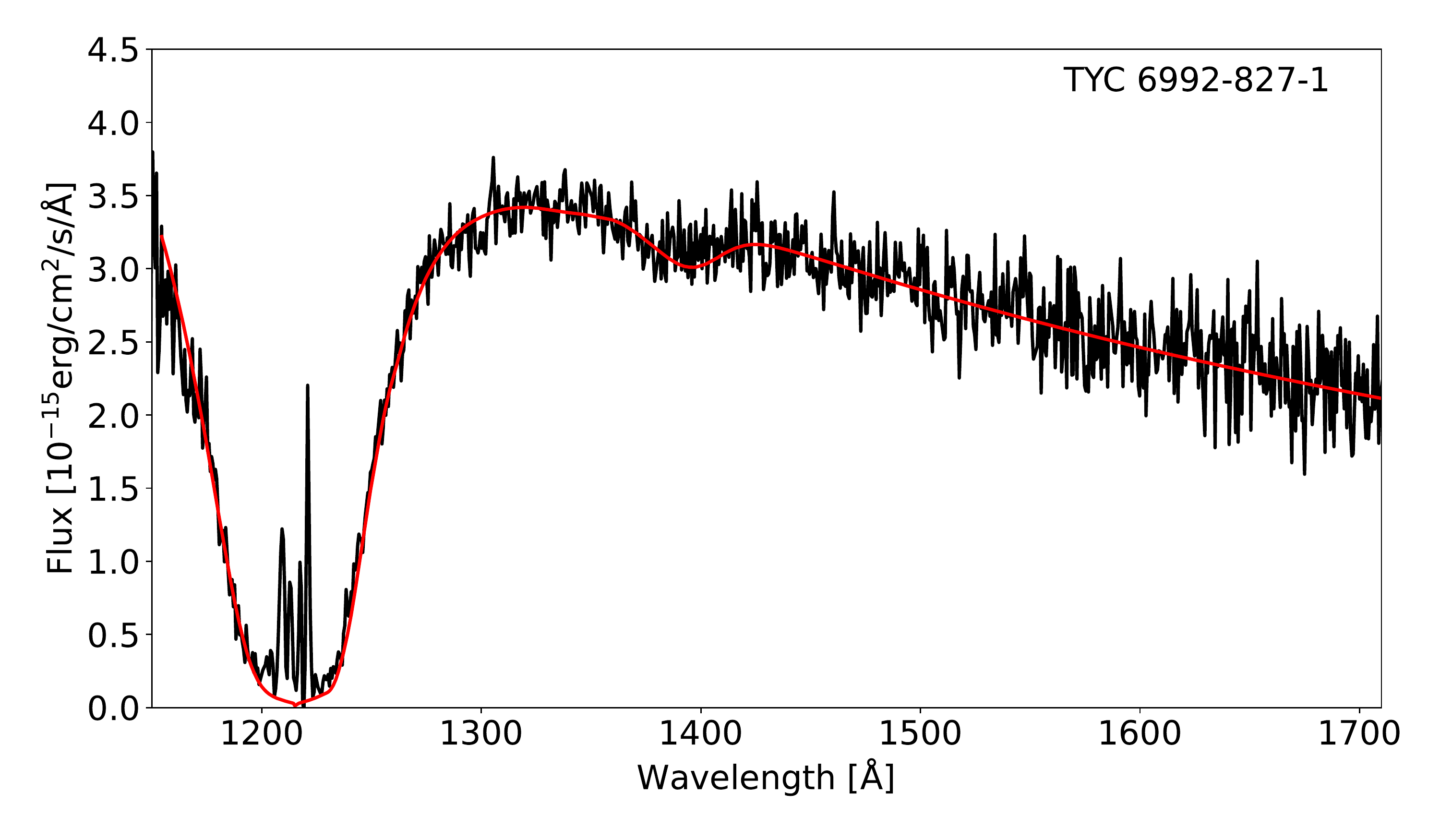}
    \includegraphics[width=\columnwidth]{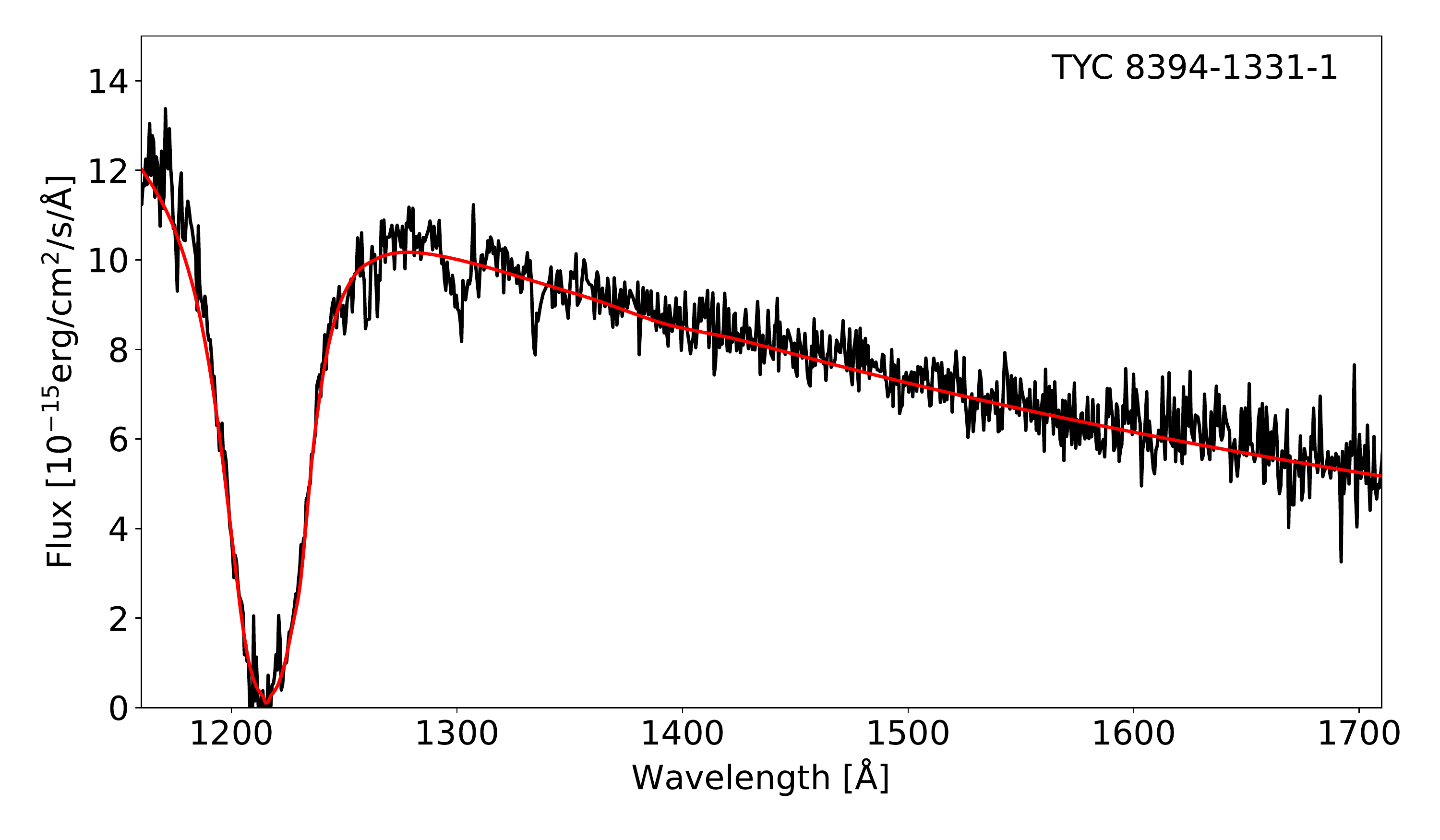}
    \caption{Fits to the {\it HST}/STIS white dwarf spectra of the three systems presented in this paper. Note that the Ly$\alpha$ geocoronal emission has not been removed, but was masked during the fitting.}
  \label{fig:hst}
  \end{center}
\end{figure}

We fitted the {\it HST}/STIS spectra to determine the effective temperatures, masses, radii and ages of the systems. To that purpose we computed a grid of synthetic white dwarf models with pure hydrogen atmospheres using the code of \citet{Koester10}. The grid spans T$_\mathrm{eff}=12\,000-30\,000$\,K (in steps of 200\,K) and $\log{g}=6.0-9.0$ (in steps of 0.1 dex). The convective mixing length parameter was set to 0.8. Thus, the free parameters were T$_\mathrm{eff}$, $\log{g}$, parallax, and reddening. Flat priors based on the edges of the grid were applied to T$_\mathrm{eff}$ and $\log{g}$, and Gaussian priors were imposed on the parallax and reddening using {\it Gaia} DR3 parallax and reddening maps, respectively. Similar to the the subgiant fits, STILISM \citep{Capitanio17} was used for 2MASS\,J1836$-$5110 and TYC\,8394$-$1331$-$1, whereas a flat prior from zero to a maximum reddening set by \citet{Schlafly11} was considered for TYC 6992$-$827$-$1. We also account for the effects of interstellar neutral hydrogen using equation 1 from \citet{Jenkins71}, however, this has a very minor effect on the final white dwarf parameters (comparable to the statistical errors) for all our targets.

The errors on the fluxes reported from STIS/MAMA observations are significantly underestimated because important noise sources were neglected, most importantly the read noise is set to zero and the gain to one\footnote{\url{https://hst-docs.stsci.edu/stisdhb/chapter-2-stis-data-structure/2-5-error-and-data-quality-array}}. Therefore, we artificially increased the errors on our spectra by including an additional ten per cent uncertainty on all the fluxes added in quadrature.

The ensemble sampler {\sc emcee} was used \citep{Foreman13}, where 100 walkers sample the parameter space during 5000 steps which ensures independent samples. In general, the autocorrelation time for each parameter was less than 100. These best-fit parameters combined with interpolation of cooling models of \citet{Althaus13} provide the masses and ages for the white dwarfs.

2MASS\,J1836$-$5110 and TYC 6992$-$827$-$1 were both observed for two {\it HST} orbits and we independently fitted the spectrum from each orbit. In the case of 2MASS\,J1836$-$5110 both spectra give consistent results and we take the average values of the two fits. However, the two spectra of TYC 6992$-$827$-$1 give significantly different results. The second spectrum is noisier and the fit is poorer. We suspect that the second spectrum may not have been centred correctly or is possibly suffering from high dark current. Therefore, for this object we report the parameters from fitting the first spectrum only. The final results for all three systems are listed in Table~\ref{tab:wdparams} and the {\it HST} spectra along with the best fitting models are shown in Figure~\ref{fig:hst}. The uncertainties reported in Table~\ref{tab:wdparams} should be considered purely statistical since they do not include any systematic uncertainties in the white dwarf models themselves. These are typically of the order of 1.5 per cent in T$_\mathrm{eff}$ and 0.04 dex in $\log{g}$ \citep[e.g.][]{Barstow03,Gianninas11}, comparable to the quoted statistical uncertainties.

Both TYC\,8394$-$1331$-$1 and 2MASS\,J1836$-$5110 show narrow absorption features in their {\it HST}/STIS spectra. These are extremely weak and narrow for 2MASS\,J1836$-$5110 and are likely interstellar in origin. However, TYC\,8394$-$1331$-$1 shows features of C\,{\sc ii} at 1335\,{\AA}, C\,{\sc iii} at 1175\,{\AA}, the silicon doublet 1260, 1265\,{\AA} and Si\,{\sc i} and Si\,{\sc iii} around 1300\,{\AA}. Several of these lines are contaminated by interstellar absorption lines, in addition the region around 1300\,{\AA} can be contaminated by geo-coronal airglow emission of O\,{\sc i}. With this caveat, we fitted the spectrum of TYC\,8394$-$1331$-$1 in order to estimate the abundances of carbon, silicon and oxygen. We computed atmospheric models which included the opacities of these elements in addition to hydrogen and compared these to the best-fit pure hydrogen model with the same temperature and surface gravity. We found a flux difference of the continuum of $<$1.5 per cent, with the largest difference at wavelengths bluer than Ly~$\alpha$. Individual abundances were also poorly constrained  due to the weakness of the features. Therefore, given the small flux difference and the large uncertainties of the abundances, we adopt the best-fit parameters from the pure hydrogen model.

With the masses of both stellar components in the binary as well as the orbital fits we can now determine the semi-major axis, $a$, using Kepler's third law:
\begin{equation}
a^3 = \frac{G(M_\mathrm{WD} + M_\mathrm{SG})P_\mathrm{orb}}{4 \pi^2},\label{eqn:kep3}
\end{equation}
where $G$ is the gravitational constant. We can also calculate the orbital inclination, $i$, via the binary mass function:
\begin{equation}
\frac{M_\mathrm{WD}^3 \sin^3{i}}{(M_\mathrm{WD} + M_\mathrm{SG})^2} = \frac{P_\mathrm{orb}K}{2 \pi G}. \label{eqn:mfunc}
\end{equation}
Both the semi-major axis and orbital inclination are listed in Table~\ref{tab:wdparams}.

\begin{table}
 \centering
  \caption{Best fit stellar parameters for the white dwarfs presented in this paper and binary parameters derived from the masses and orbital fits.}
  \label{tab:wdparams}
  \tabcolsep=0.05cm
  \begin{tabular}{@{}lccc@{}}
    \hline
    Parameter & 2MASS\,J1836$-$5110 & TYC\,6992$-$827$-$1 & TYC\,8394$-$1331$-$1 \\
    \hline
    \multicolumn{4}{c}{From spectral fit:} \\
    T$_\mathrm{eff,WD}$ [K]     & $22250\pm250$ & $15750\pm50$ & $19400\pm100$ \\
    $\log{g}_\mathrm{WD}$ [dex] & $7.49\pm0.03$ & $7.14\pm0.02$ & $6.53\pm0.03$ \\
    \multicolumn{4}{c}{Derived:} \\
    M$_\mathrm{WD}$ [\MSUN]     & $0.40\pm0.01$ & $0.28\pm0.01$ & $0.24\pm0.01$ \\
    $\tau_\mathrm{cool,WD}$ [Myr]  & $7.2\pm0.1$ & $6.9\pm0.1$  & $175\pm21$\\
    $a$ [au]  & $1.42\pm0.04$ & $0.27\pm0.01$ & $0.31\pm0.01$ \\
    $i$ [deg] & $47\pm4$      & $26\pm2$      & $39\pm2$      \\
    \hline
  \end{tabular}
\end{table}

\section{Notes on individual systems}

\begin{figure}
  \begin{center}
    \includegraphics[width=\columnwidth]{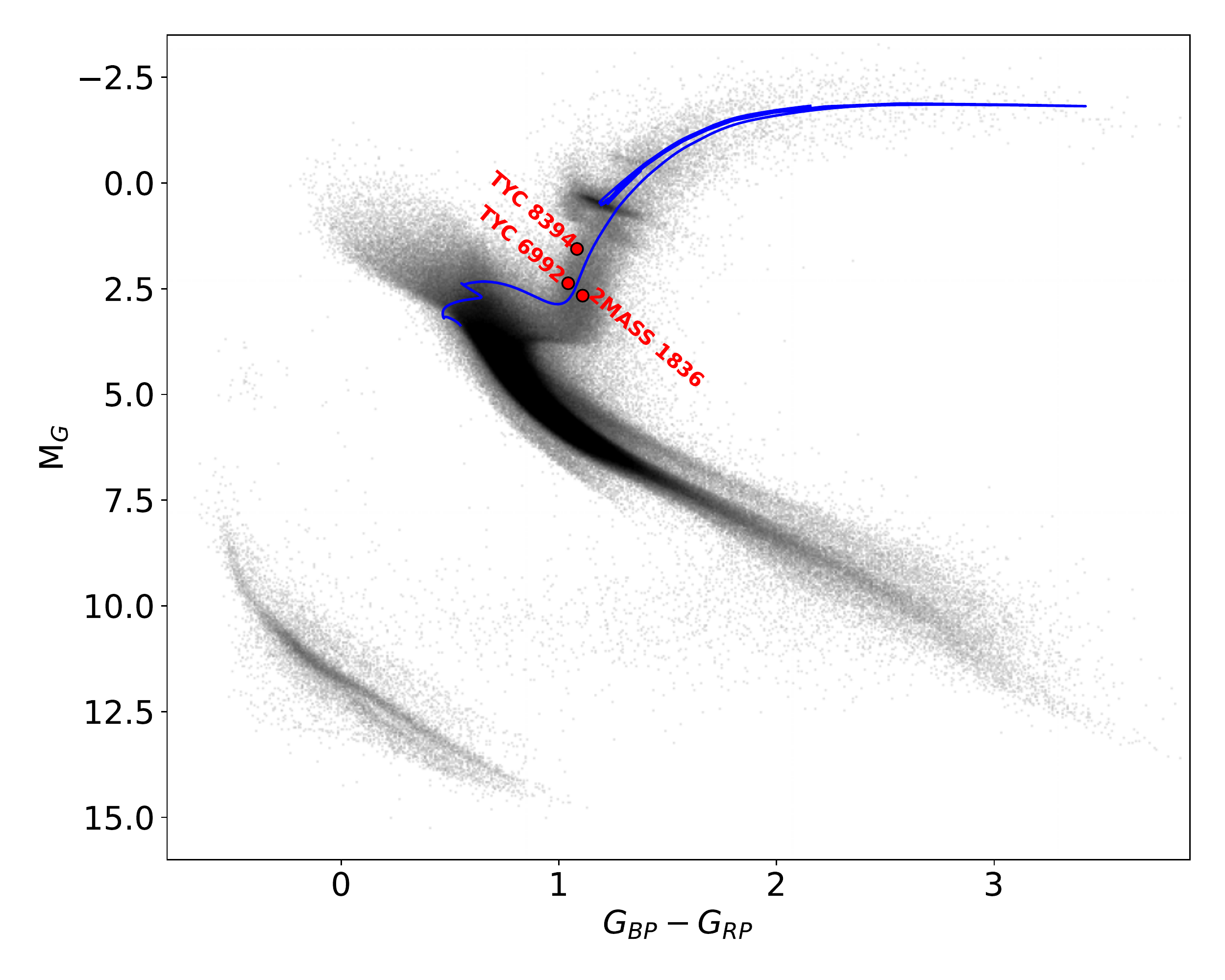}
    \caption{{\it Gaia} DR3 Hertzsprung-Russell diagram with the location of the three systems presented in this paper indicated in red. The blue line shows a 1.3{\MSUN} MIST evolutionary track. A random sample of objects within 500\,pc is shown in black.}
  \label{fig:hrd}
  \end{center}
\end{figure}

In this section we describe each system in detail. The locations of all three systems in the {\it Gaia} Hertzsprung-Russell diagram are shown in Figure~\ref{fig:hrd}, along with an 1.3{\MSUN} evolutionary track for reference, taken from the Modules for Experiments in Stellar Astrophysics (MESA; \citealt{Paxton11}) Isochrones and Stellar Tracks (MIST) model grids \citep{Dotter16,Choi16}.

\subsection{2MASS\,J1836$-$5110}

With an orbital period of 461 days, 2MASS\,J1836$-$5110 has the longest period of the three systems presented in this paper and is comparable to the orbital periods of the {\it Kepler} self-lensing systems. Despite the low eccentricity ($e=0.028\pm0.001$), the probability of falsely rejecting a circular fit is low according to the proposal of \citet{Lucy71}, hence we consider the measured eccentricity to be accurate and the binary is not quite circular. While the white dwarf mass is the highest of our three systems ($0.40\pm0.01$\,\MSUN), it is still much lower than typically seen in detached white dwarf binaries \citep[e.g.][]{Zorotovic11}, including the self-lensing systems (with the exception of the very low mass white dwarf in KIC 8145411, \citealt{Masuda19}).

We note that 2MASS\,J1836$-$5110 has a slightly high renormalised unit weight error (RUWE) of 1.6 in {\it Gaia} DR3 \citep{Gaia21}, which is often indicative of binarity. Given the long orbital period of this system it is likely that the binary motion causes some astrometric variability, leading to this high value (the semi-major axis for this system is $1.42\pm0.04$\,au). However, this system does not feature in any of the {\it Gaia} DR3 non-single star catalogues \citep{Gaia22}. This is likely due to its relatively large distance of 900\,pc and the low mass of the white dwarf, although the orbital inclination of only $47\pm4$ degrees is favourable for astrometric detection. There are no clear signs that this binary is part of a higher order system.

\subsection{TYC\,6992$-$827$-$1}

The orbital period of TYC\,6992$-$827$-$1 is 41 days, placing it between the short period (post-common envelope) systems and longer period self-lensing systems. To date, all of the ultraviolet excess objects from \citet{Parsons16} with periods in this range have turned out to be contaminants, from either active stars in binaries or triple systems, where the ultraviolet excess arises from a distant white dwarf companion to a main-sequence binary \citep{Lagos20,Lagos22}. However, in this case we can rule out stellar activity, since our {\it HST} data clearly reveal that the ultraviolet excess is caused by a white dwarf. Moreover, the extremely low mass for the white dwarf ($0.28\pm0.01$\,{\MSUN}) must be the result of binary interaction, since a single star could not produce such a low mass white dwarf within a Hubble time, making it unlikely that the white dwarf is a distant companion to a binary. Finally, we note that the contaminating main-sequence binaries identified in \citet{Lagos20,Lagos22} all have high eccentricities ($e>0.26$), which is not the case for TYC\,6992$-$827$-$1. While our fit to the radial velocities does favour a very small eccentricity, we cannot reject the possibility of a circular orbit in this binary \citep{Lucy71}, in stark contrast to the \citet{Lagos20,Lagos22} systems.

Similar to 2MASS\,J1836$-$5110, TYC\,6992$-$827$-$1 also has a very high RUWE value in {\it Gaia} DR3 of 4.2 \citep{Gaia21}. This system is somewhat closer at 500\,pc, but the much shorter orbital period and lower stellar masses makes the astrometric signal from this binary weaker (the semi-major axis for this system is $0.27\pm0.01$\,au), although the low inclination of this system ($26\pm2$ degrees) will somewhat boost the astrometric signal. Therefore, while it is likely that the binary itself is responsible for the high RUWE value, it is possible that TYC\,6992$-$827$-$1 is part of a higher order system, with a distant companion adding to the high astrometric scatter. The system is listed as a single-lined spectroscopic binary in {\it Gaia} DR3 \citep{Gaia22}, with orbital parameters consistent with our fit (P$_\mathrm{orb}=41.32\pm0.06$\,d, $e=0.09\pm0.07$, $K=5.5\pm0.3$\,\kms) , but is not classified as an astrometric binary, hence we cannot confirm if it is a higher order system.

\subsection{TYC\,8394$-$1331$-$1}

TYC\,8394$-$1331$-$1 stands out from the other two objects because it is clearly a triple system. However, based on the radial velocity data we can confidently conclude that the white dwarf must be part of the inner 52 day binary. This is because the mass function implies a minimum companion mass of 0.14\,{\MSUN} for the inner orbit and 0.39\,{\MSUN} for the outer orbit. We know that the white dwarf has a mass of $0.24\pm0.01$\,{\MSUN} from our {\it HST} spectrum, therefore, it cannot be the outer companion, since this would require an non-physical inclination (i.e. the measured white dwarf mass and outer orbit parameters give $\sin{i}>1$ in Equation~\ref{eqn:mfunc}). Using the inner orbit parameters gives an inclination for the inner binary of $39\pm2$ degrees. If the outer orbit is co-planar with this then the mass of the outer companion is $0.62\pm0.04$\,{\MSUN}. If this outer companion is a main-sequence star then it can easily be hidden in the glare of the luminous subgiant star, which would be consistent with the lack of any clear infrared excess in the SED. 

The inner binary has a semi-major axis of $0.31\pm0.01$\,au, while the outer orbit has a semi-major axis of 2.1-2.2\,au, depending upon the mutual inclination of the orbits. Despite being a compact triple ($a_\mathrm{out}/a_\mathrm{in}\simeq7$) the system is secularly stable in its current configuration, due mainly to the very circular orbits \citep{Toonen20}. Indeed, given the uncertainties on our measurements we cannot exclude the possibility that both the inner and outer orbits are circular \citep{Lucy71}. Our measured parameters place the system outside of eccentric Lidov-Kozai (LK) regime (in which extreme eccentricities close to one can be achieved), but it will still experience lower order LK cycles \citep[i.e. quadrupole order][]{Lidov62,Kozai62,Naoz16}. Depending upon the mutual inclination, this could push the eccentricities as high as 0.77. 

It is worth noting that mass loss from the central binary, due to the evolution of the white dwarf progenitor, is likely to have increased the size of the outer orbit in the past. The original system would therefore have been even more compact than at present and hence the LK timescales would have been much shorter in the past. It is possible that eccentricity oscillations could have played a role in triggering the mass transfer phase in the inner binary before the white dwarf progenitor had a chance to evolve, thus forming the very low mass white dwarf we see today \citep[e.g.][]{Toonen20}.

TYC\,8394$-$1331$-$1 was identified as a non-single source in {\it Gaia} DR3 with clear astrometric acceleration, but no period was measured \citep{Gaia22}. This is likely due to the compact triple nature of the system, since such high order orbits were not modelled in {\it Gaia} DR3. The period of the outer orbit is also at the upper end of what can be detected with the baseline of {\it Gaia} DR3.

\section{Discussion}

\begin{figure*}
  \begin{center}
    \includegraphics[width=\textwidth]{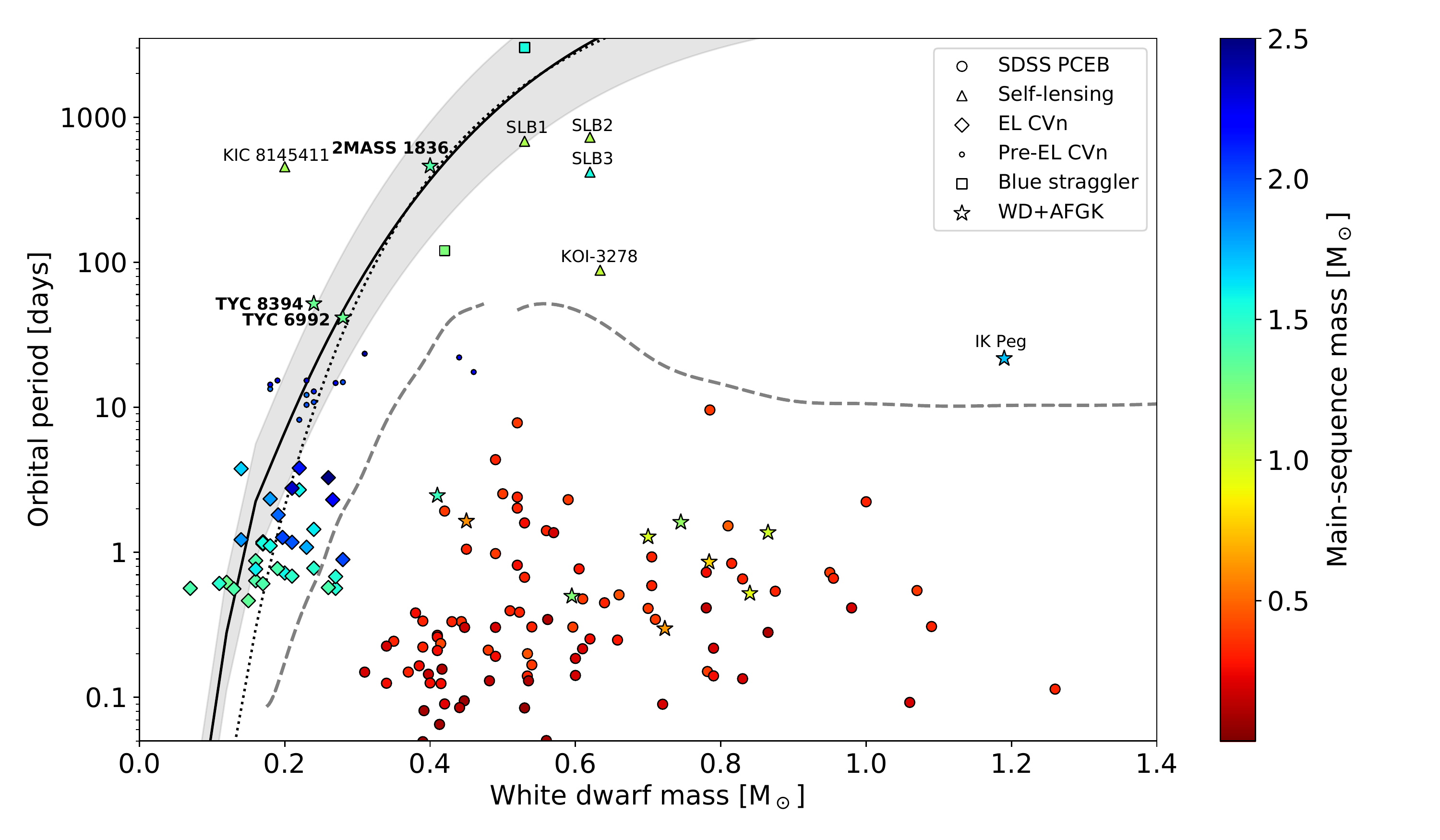}
    \caption{Orbital period and mass distributions for post-mass transfer white dwarf binaries. The grey-dashed line is the maximum orbital period for a post-common envelope system with a 1{\MSUN} companion assuming high efficiency (i.e. $\alpha_\mathrm{CE} = 1$), but no additional energy sources \citep{Zorotovic14}. The black line and grey shaded region is the theoretical $P_\mathrm{orb}-M_\mathrm{WD}$ relation for stable mass transfer \citep{Rappaport95} while the black dotted line a modified version of this for shorter period systems \citep{Lin11}. Large circles show post-common envelope binaries found in the Sloan Digital Sky Survey (SDSS) from \citet{Rebassa10,Rebassa16}. Triangles show self-lensing systems \citep{Kruse14,Kawahara18,Masuda19}. Diamonds are EL~CVn systems from \citet{Maxted14} and \citet{Roestel18}. Small circles are the EL~CVn progenitor systems from \citet{ElBadry22}. Squares are blue stragglers from \citet{Gosnell19}. Star symbols show WD+AFGK systems \citep{Landsman93,OBrien01,Parsons15,Krushinsky20,Hernandez21,Hernandez22}, including the three presented in this paper, which are highlighted in bold.   
    }
  \label{fig:pvm}
  \end{center}
\end{figure*}

With precise stellar and binary parameters in hand for all three systems, in this section we reconstruct their past evolution and predict their future fates. For completeness, we include a list of the stellar and binary parameters for all of the systems discovered in The White Dwarf Binary Pathways Survey to date (include the three presented in this paper) in Table~\ref{tab:wd+afgk} in the Appendix.

\subsection{Past evolution}

In Figure~\ref{fig:pvm} we show the orbital period as a function of white dwarf mass for the observed sample of close white dwarf binaries with non-degenerate companions. The mass of the companion is also shown colour coded. The dashed line is the maximum orbital period for post-common envelope systems assuming a 1{\MSUN} companion and a highly efficient use of the orbital energy during the common-envelope phase, i.e. $\alpha_\mathrm{CE} = 1$, but with no contribution of any other energy sources \citep{Zorotovic14}. The three systems studied in this paper lie well above this line, which gives us a first indication that they are most likely not the result of a common envelope phase. We note that there are two systems above this line, KOI\,3278 and IK\,Peg, for which common envelope evolution remains the most likely explanation for their current orbital configuration \citep{Zorotovic14}. However, both contain more massive white dwarfs ($>0.5$\,{\MSUN}) that descend from progenitors on the asymptotic giant branch, i.e. with extended envelopes, and therefore the inclusion of a very small fraction of recombination energy in the binding energy of the envelope is sufficient to explain their current orbital periods. Given the low mass of the white dwarfs in the three systems studied here, which imply their progenitors were not highly evolved, it does not seem likely that the inclusion of recombination energy in the calculation of the binding energy will help us to explain their past evolution assuming a common envelope phase. 

In order to definitively rule out that these are post-common envelope binaries, we decided to reconstruct the common envelope phase using the same algorithm described in detail in \citet{Hernandez21}, based on the binary star evolution (BSE) code from \citet{Hurley2002}. As we expected, no possible progenitors were found for any of the three systems, even assuming the most efficient use of orbital energy during the common-envelope phase ($\alpha_\mathrm{CE} = 1$). We then also allowed our reconstruction algorithm to include a variable fraction of the hydrogen recombination energy that contributed to expel the envelope ($\alpha_\mathrm{rec}$ in \citealt{Zorotovic14}). Even in the unrealistic case of including 100 per cent of this energy in the calculations of the envelope's binding energy, i.e. $\alpha_\mathrm{CE} = 1$ and $\alpha_\mathrm{rec} = 1$, we were not able to find any possible progenitors. We therefore conclude that the three systems most likely experienced a phase of dynamically stable but non-conservative mass transfer, which is consistent with their location in Figure~\ref{fig:pvm} close to the theoretical  $P_\mathrm{orb}-M_\mathrm{WD}$ relation for stable mass transfer from \citet{Rappaport95}. 

We also include EL~CVn binaries (close pre-white dwarfs with A/F main-sequence star companions) in Figure~\ref{fig:pvm}, which all sit close to the theoretical $P_\mathrm{orb}-M_\mathrm{WD}$ relation for stable mass transfer \citep{Maxted14,Roestel18}. The location of these systems is unsurprising given the very low (pre-) white dwarf masses in these systems, which imply that mass transfer occurred when the originally more massive star was at the end of the main-sequence or shortly afterwards on the subgiant branch \citep{Chen17}. For EL~CVn systems the mass transfer is stable and a common envelope is avoided. It is thought that the mass transfer is at least partially conservative for EL CVn systems, given the high masses of the companion stars \citep{Chen17}. Given the lower companion masses in our new long period systems the mass transfer was likely far less conservative in these cases (if at all), but the overall evolution may well be quite similar. In particular, \citet{Lagos20b} showed that EL~CVn systems are generally the inner binaries of hierarchical triples. This is because of the need for the original main-sequence binary to be close in order to start mass transfer at the end of the main-sequence and virtually all known close main-sequence binaries ($P_\mathrm{orb}<3$\,days) are  known to be the inner binaries of hierarchical triple systems \citep{Tokovinin06}. We know that TYC\,8394$-$1331$-$1 is also the inner binary of a hierarchical triple and it is possible that TYC\,6992$-$827$-$1 is as well, hence there are clear similarities between our systems and EL~CVn binaries. 

We also include the recently discovered EL~CVn progenitor systems identified by \citet{ElBadry22}. These single-lined binaries were originally though to be high-mass function systems, but were shown instead to be binaries containing highly stripped low mass giant donors, the majority of which are recently detached systems (although at least one system appears to still be mass transferring). In figure~\ref{fig:pvm} these systems sit between the EL~CVn binaries and our newly discovered systems, reinforcing the link between these populations.

We also point out the similarity of our systems to the long period hot subdwarf-B (sdB) binaries presented in \citet{Vos19}, which are thought to be the result of stable non-conservative mass transfer. Our systems appear to represent the extreme lower white dwarf/sdB mass and orbital period limit of this sample and a possible link between these systems and EL\,CVn binaries. Our systems also closely resemble blue stragglers \citep[e.g.][]{Carney01,Gosnell19,Leiner19}, which are also thought to descend from stable mass transfer, although they often have moderately high eccentricities, in contrast to our systems. Two blue straggler systems with spectroscopically confirmed white dwarf companions are also shown in Figure~\ref{fig:pvm}.

Until now, all of the WD+AFGK systems characterised by our survey have stellar and binary parameters consistent with being typical post-common envelope binaries that do not require a highly efficient envelope removal \citep{Parsons15,Hernandez21,Hernandez22}, which is also the case for all white dwarf plus M dwarf or brown dwarf binaries \citep{Zorotovic22}. All of the longer period systems have either been contaminants \citep{Lagos20,Lagos22} or the descendants of stable mass transfer. This implies that $\alpha_\mathrm{CE}$ may always be small and independent of the companion mass. Long period post-common envelope systems (e.g. IK\,Peg or KOI\,3278) are likely to be rare.

It is clear that the WD+AFGK systems consistent with common envelope evolution (i.e. those with short orbital periods and more typical white dwarf masses) descend from binaries that were originally wider, allowing the white dwarf progenitor to evolve before mass transfer began. Hence the original binary separation plays a key role in determining the type of mass transfer that occurs and the nature of the resulting system. The original mass ratio is also important, since large mass ratios exclude stable mass transfer, hence the reason why all the close white dwarf plus M dwarf binaries are clearly post-common envelope systems \citep{Ge22}. SLB1, 2 and 3 all likely had original orbits similar to the post-common envelope systems, but mass ratios close to one and hence evolved through stable mass transfer to their current long orbits. Therefore, to form the systems presented in this paper requires an initially short period and an original mass ratio not too different from one.

\subsection{Future evolution}

We also use the BSE code in order to predict the future evolution of the three systems. According to the current orbital period and mass ratio, the three will enter a common envelope phase when the subgiant companion evolves to the first giant branch. Using our standard small efficiency ($\alpha_\mathrm{CE} = 0.2$-0.3) and without recombination energy TYC\,6992$-$827$-$1 and TYC\,8394$-$1331$-$1 are expected to experience a merger during the common-envelope phase (note that we did not include any influence from the outer companion when simulating the future of TYC\,8394$-$1331$-$1). What the outcome of these mergers will be is not clear, but these two systems will most likely end up as single white dwarfs, somewhat more massive than the low-mass white dwarfs we see today. This may be a possible formation channel for creating single low mass white dwarfs \citep{Nelemans98}. On the other hand, the longer orbital period derived for 2MASS\,J1836$-$5110 implies it will most likely survive the common envelope phase and become a close double helium-core white dwarf system. The later is expected to emerge from the common envelope phase with white dwarf masses of $\sim$0.40{\MSUN} (the white dwarf we observe today) and $\sim$0.45{\MSUN} at an orbital period of 0.37-0.67 days. A number of close double white dwarf systems containing two low mass white dwarfs are already known (e.g. \citealt{Bours14,Parsons20}, see \citealt{Schreiber22} for a full list) and these binaries may well have descended from similar systems to those presented in this paper. Indeed \citet{Nelemans20} noted that binaries containing two low mass white dwarfs are difficult to reproduce via two common envelope events.

All three systems might also appear as symbiotic systems for a short period ($\sim$100$-$200\,Myr) before entering the common envelope phase, while the giants are still under-filling their Roche Lobes and mass transfer comes from a wind. 

Finally, we reiterate the fact that the cooling ages of the white dwarfs in both 2MASS\,J1836$-$5110 and TYC\,6992$-$827$-$1 are comparable (or even shorter than) the thermal timescale of the subgiant stars and hence the stellar parameters for the subgiants in these two systems may not be completely accurate. While this does not affect our conclusions of the past evolution of these systems (where the orbital period and white dwarf mass are far more important), it may alter the future evolution of these systems from our predictions if the companion stars have lower masses than our estimates. In particular, if we have strongly overestimated their masses then it is possible that these systems will not evolve for longer than a Hubble time and hence would not be representative of current day double white dwarf systems. Measuring the orbits astrometrically would place strong additional constraints on the masses and hence future {\it Gaia} data releases may well resolve this issue.

\section{Conclusions}

We have discovered three low mass white dwarfs in binaries with subgiant stars with orbital periods substantially longer than typical post-common envelope binaries. Using both ground- and space-based spectroscopy we placed precise constraints on the stellar and binary parameters of all three systems, revealing one to be the inner binary of a hierarchical triple. The combination of long orbital periods and low white dwarf masses mean that we are unable to reproduce these systems via common envelope evolution, even assuming highly efficient envelope ejection and the addition of recombination energy. We therefore conclude that these systems must be the result of stable but non-conservative mass transfer, which demonstrates that white dwarfs in binaries with intermediate mass A, F, G and early-K companions can follow radically different evolutionary pathways depending upon the initial separation and mass ratio of the binary.

\section*{Acknowledgements}

SGP acknowledges the support of a Science and Technology Facilities Council (STFC) Ernest Rutherford Fellowship. MSH, OT, and MRS acknowledge support by ANID -- Millennium Science Initiative Program -- NCN19\_171. OT was also supported by FONDECYT (grant 321038). MRS and MZ were also supported by FONDECYT (grant 1221059). BTG was supported by grant ST/T000406/1 from the STFC. This project has received funding from the European Research Council (ERC) under the European Union’s Horizon 2020 research and innovation programme (Grant agreement No. 101020057). RR has received funding from the postdoctoral fellowship programme Beatriu de Pin\'os, funded by the Secretary of Universities and Research (Government of Catalonia) and by the Horizon 2020 programme of research and innovation of the European Union under the Maria Sk\l{}odowska-Curie grant agreement No 801370. ARM acknowledges support from Grant RYC$-$2016$-$20254 funded by MCIN/AEI/10.13039/501100011033 and by ESF Investing in your future, and from MINECO under the PID2020$-$117252GB$-$I00 grant. The results presented in this paper are based on observations collected at the European Southern Observatory under programme IDs 096.D-0217, 098.D-0023 and 108.D-0122. For the purpose of open access, the author has applied a creative commons attribution (CC BY) licence to any author accepted manuscript version arising.

\section*{Data Availability Statement}

Raw and reduced FEROS, UVES and X-shooter data are available through the ESO archive at \url{http://archive.eso.org/cms.html}. CHIRON and Du Pont data will be shared on reasonable request to the corresponding author.

\bibliographystyle{mnras}
\bibliography{longp}

\appendix

\section{Radial velocity measurements}

\begin{table*}
 \centering
  \caption{Velocity measurements for 2MASS\,J1836$-$5110}
  \label{tab:2mass1836_vels}
  \begin{tabular}{@{}lccc@{}}
    \hline
    BJD (mid-exposure) & RV (\kms) & Err (\kms) & Instrument \\
    \hline
2456809.76248753 & 26.235 & 0.500 & Du Pont \\
2456810.75213144 & 26.323 & 0.500 & Du Pont \\
2456819.74917433 & 26.756 & 0.200 & FEROS \\
2456819.76019754 & 26.716 & 0.200 & FEROS \\
2456819.73815273 & 26.821 & 0.200 & FEROS \\
2456827.75170523 & 27.362 & 0.013 & FEROS \\
2456828.62383469 & 27.431 & 0.010 & FEROS \\
2456835.60862554 & 27.925 & 0.010 & FEROS \\
2457189.68636435 & 20.912 & 0.010 & FEROS \\
2457472.81628360 & 27.562 & 0.010 & FEROS \\
2457587.66341401 & 20.571 & 0.010 & FEROS \\
2458290.66927933 & 31.068 & 0.017 & FEROS \\
2458290.75708253 & 31.052 & 0.020 & FEROS \\
2458290.83118133 & 31.096 & 0.021 & FEROS \\
2458291.72839868 & 31.092 & 0.015 & FEROS \\
2458292.72061165 & 31.032 & 0.014 & FEROS \\
2458292.89637992 & 31.274 & 0.023 & FEROS \\
2458293.69924433 & 31.052 & 0.020 & FEROS \\
2458293.80774890 & 31.105 & 0.028 & FEROS \\
2458600.70033571 & 22.235 & 0.025 & FEROS \\
2458601.72229306 & 22.333 & 0.022 & FEROS \\
2458602.85311495 & 22.350 & 0.021 & FEROS \\
2458602.88968531 & 22.255 & 0.019 & FEROS \\
2458603.77265516 & 22.477 & 0.022 & FEROS \\
2458603.83193249 & 22.378 & 0.024 & FEROS \\
2458606.77059273 & 22.657 & 0.019 & FEROS \\
2458606.77465320 & 22.578 & 0.020 & FEROS \\
2458606.77871538 & 22.644 & 0.018 & FEROS \\
2458606.78279148 & 22.569 & 0.018 & FEROS \\
2458606.78686883 & 22.663 & 0.019 & FEROS \\
2459530.52684574 & 22.539 & 0.500 & UVES \\
2459508.59217903 & 21.553 & 0.500 & UVES \\
2459647.88459790 & 30.384 & 0.500 & UVES \\
2459653.77530590 & 30.416 & 0.500 & UVES \\
    \hline
  \end{tabular}
\end{table*}

\begin{table*}
 \centering
  \caption{Velocity measurements for TYC\,6992$-$827$-$1}
  \label{tab:tyc6992_vels}
  \begin{tabular}{@{}lccc@{}}
    \hline
    BJD (mid-exposure) & RV (\kms) & Err (\kms) & Instrument \\
    \hline
2456810.88987164 & -45.535 & 0.500 & Du Pont \\
2456828.77035696 & -38.964 & 0.010 & FEROS \\
2456829.81668064 & -38.271 & 0.012 & FEROS \\
2456829.91031794 & -38.255 & 0.010 & FEROS \\
2456830.81378745 & -37.772 & 0.010 & FEROS \\
2456831.91639108 & -37.303 & 0.010 & FEROS \\
2457002.63027686 & -36.798 & 0.010 & FEROS \\
2457003.63339269 & -37.072 & 0.010 & FEROS \\
2457004.54995757 & -37.425 & 0.010 & FEROS \\
2457025.59357472 & -46.098 & 0.500 & Du Pont \\
2457026.53553894 & -46.043 & 0.500 & Du Pont \\
2457027.55362046 & -45.898 & 0.500 & Du Pont \\
2457188.82273257 & -47.183 & 0.010 & FEROS \\
2457250.71244946 & -36.994 & 0.500 & CHIRON \\
2457251.83661388 & -37.380 & 0.500 & CHIRON \\
2457252.79154637 & -37.701 & 0.500 & CHIRON \\
2457261.66273899 & -43.801 & 0.500 & CHIRON \\
2457269.64503261 & -47.782 & 0.500 & CHIRON \\
2457283.71222538 & -39.627 & 0.500 & CHIRON \\
2457294.66224794 & -37.917 & 0.500 & CHIRON \\
2457297.69522221 & -39.590 & 0.500 & CHIRON \\
2457303.73600960 & -44.364 & 0.500 & CHIRON \\
2457311.68670972 & -47.468 & 0.500 & CHIRON \\
2457313.63664922 & -47.656 & 0.500 & CHIRON \\
2457319.61130834 & -43.733 & 0.500 & CHIRON \\
2457329.61830650 & -36.986 & 0.500 & UVES \\
2457332.60726505 & -36.971 & 0.500 & CHIRON \\
2457340.53546990 & -40.643 & 0.500 & UVES \\
2457364.59907431 & -41.053 & 0.500 & CHIRON \\
    \hline
  \end{tabular}
\end{table*}

\begin{table*}
 \centering
  \caption{Velocity measurements for TYC\,8394$-$1331$-$1}
  \label{tab:tyc8394_vels}
  \begin{tabular}{@{}lccc@{}}
    \hline
    BJD (mid-exposure) & RV (\kms) & Err (\kms) & Instrument \\
    \hline
2456873.776669648 & 39.523 & 0.500 & CHIRON \\
2456888.587295888 & 30.472 & 0.500 & CHIRON \\
2456906.592794124 & 27.957 & 0.500 & CHIRON \\
2456911.562011322 & 31.399 & 0.500 & CHIRON \\
2456920.612289260 & 36.885 & 0.500 & CHIRON \\
2456924.559734249 & 37.633 & 0.500 & CHIRON \\
2456926.577372765 & 37.477 & 0.500 & CHIRON \\
2456931.537606544 & 35.435 & 0.500 & CHIRON \\
2456937.555521341 & 31.019 & 0.500 & CHIRON \\
2456941.553596461 & 27.688 & 0.500 & CHIRON \\
2456944.514096235 & 25.927 & 0.500 & CHIRON \\
2457189.715487490 & 37.899 & 0.010 & FEROS \\
2457222.548450931 & 33.548 & 0.500 & CHIRON \\
2457223.670184840 & 34.324 & 0.500 & CHIRON \\
2457224.662427607 & 35.183 & 0.500 & CHIRON \\
2457249.605290546 & 35.341 & 0.500 & CHIRON \\
2457250.611555887 & 33.873 & 0.500 & CHIRON \\
2457251.743665238 & 33.247 & 0.500 & CHIRON \\
2457252.701218139 & 32.589 & 0.500 & CHIRON \\
2457257.570658996 & 30.039 & 0.500 & CHIRON \\
2457267.494933716 & 31.393 & 0.500 & CHIRON \\
2457273.705291479 & 35.463 & 0.500 & CHIRON \\
2457277.662845165 & 38.429 & 0.500 & CHIRON \\
2457283.616742769 & 42.241 & 0.500 & CHIRON \\
2457293.603606648 & 41.915 & 0.500 & CHIRON \\
2457319.564909399 & 33.068 & 0.500 & CHIRON \\
2457333.536274699 & 43.315 & 0.500 & CHIRON \\
2457454.882325782 & 42.476 & 0.500 & UVES \\
2457635.574287844 & 35.884 & 0.500 & UVES \\
2457640.602255318 & 39.406 & 0.500 & UVES \\
2457645.614058001 & 42.386 & 0.500 & UVES \\
2457665.541105256 & 34.940 & 0.500 & UVES \\
2457687.515129791 & 33.677 & 0.500 & UVES \\
2459371.918669399 & 40.037 & 0.550 & X-SHOOTER \\
2459508.581898705 & 34.272 & 0.500 & UVES \\
2459530.539642073 & 30.267 & 0.500 & UVES \\
2459538.528942762 & 24.979 & 0.500 & UVES \\
2459541.542958050 & 24.058 & 0.500 & UVES \\
2459649.875344526 & 23.428 & 0.500 & UVES \\
2459651.892380544 & 24.322 & 0.500 & UVES \\
2459668.886111152 & 35.160 & 0.500 & UVES \\
2459672.809120230 & 36.119 & 0.500 & UVES \\
2459673.828855139 & 36.316 & 0.500 & UVES \\
    \hline
  \end{tabular}
\end{table*}

\section{SED fit corner plots}

\begin{figure*}
  \begin{center}
  	\vspace{50mm}
    \includegraphics[width=0.33\textwidth]{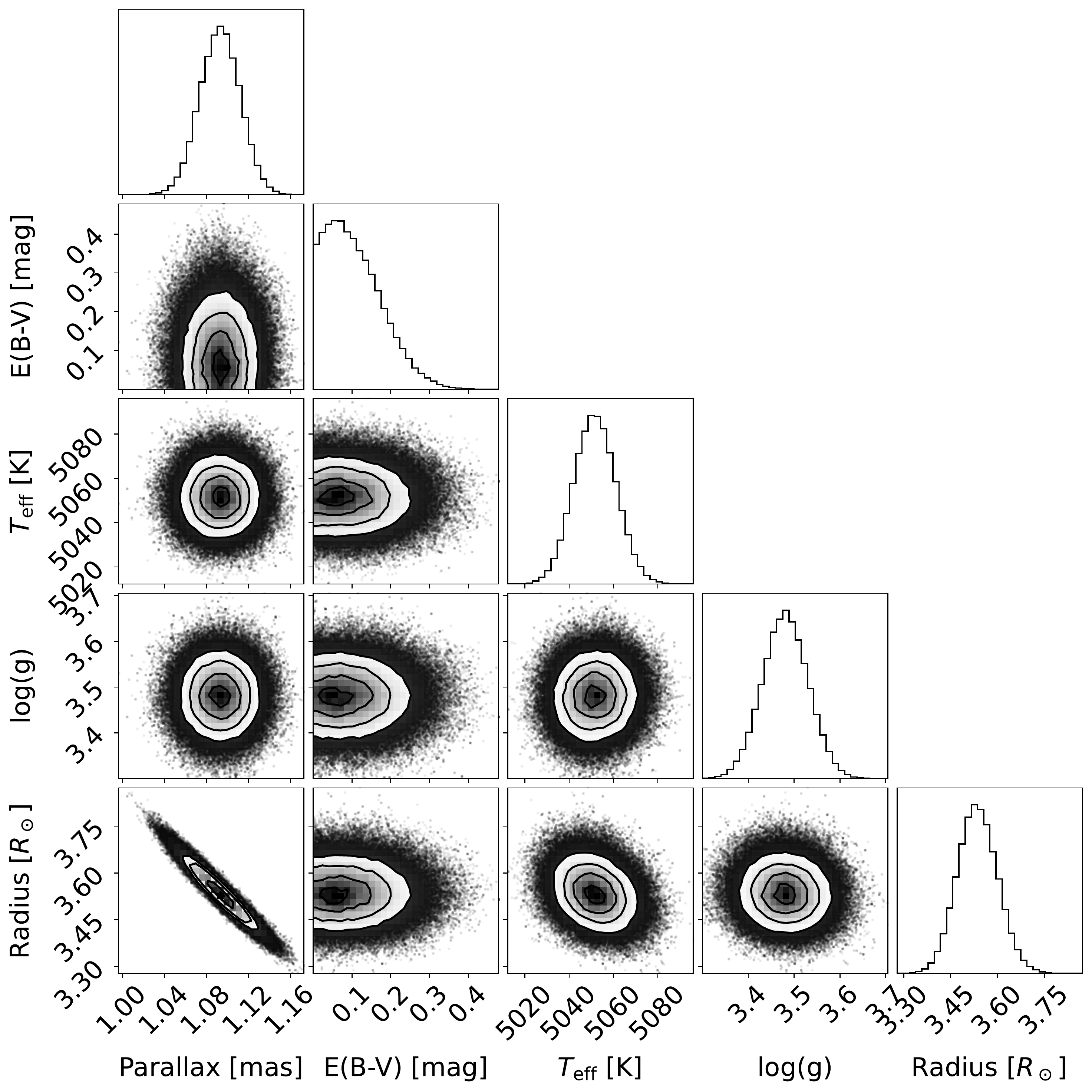}
    \includegraphics[width=0.33\textwidth]{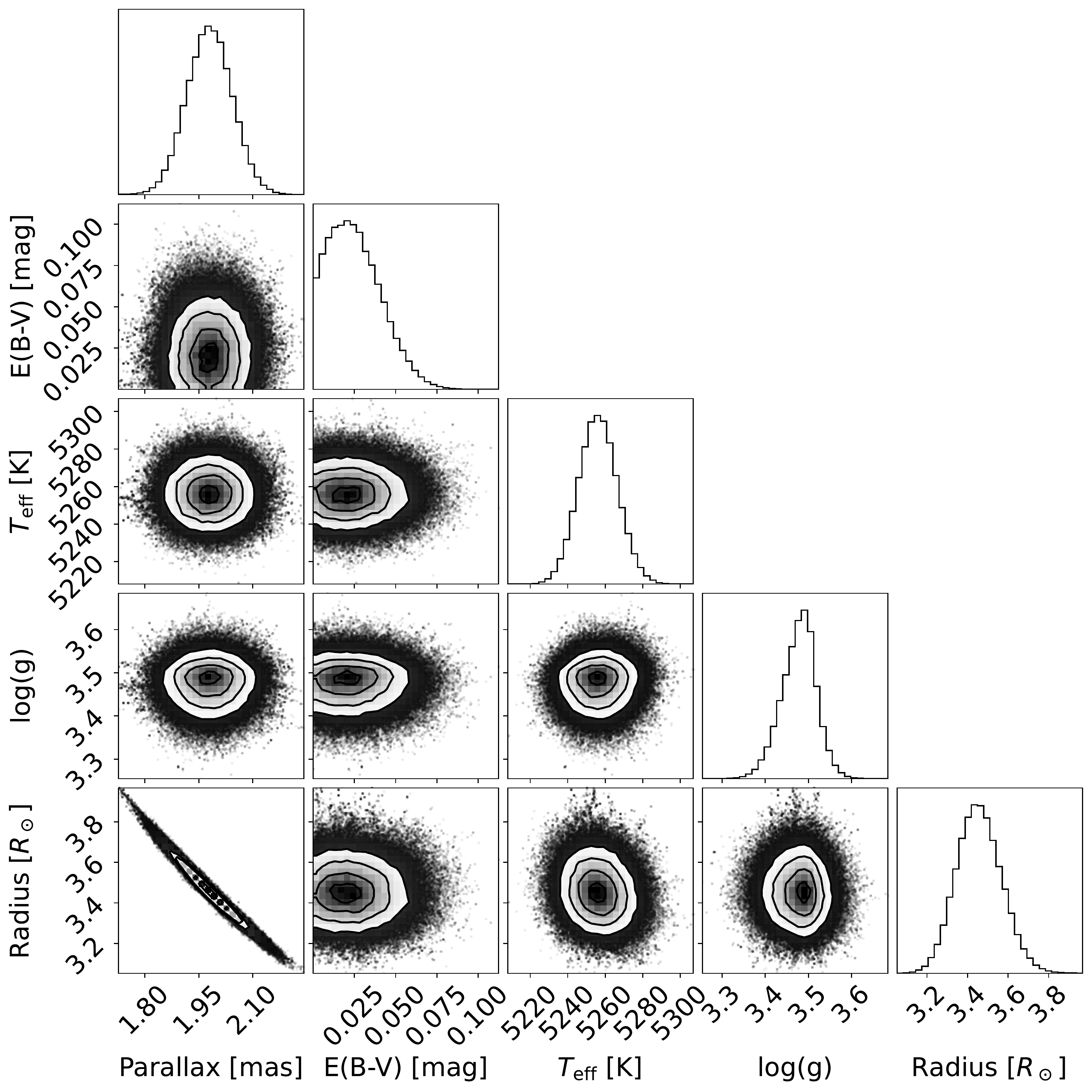}
    \includegraphics[width=0.33\textwidth]{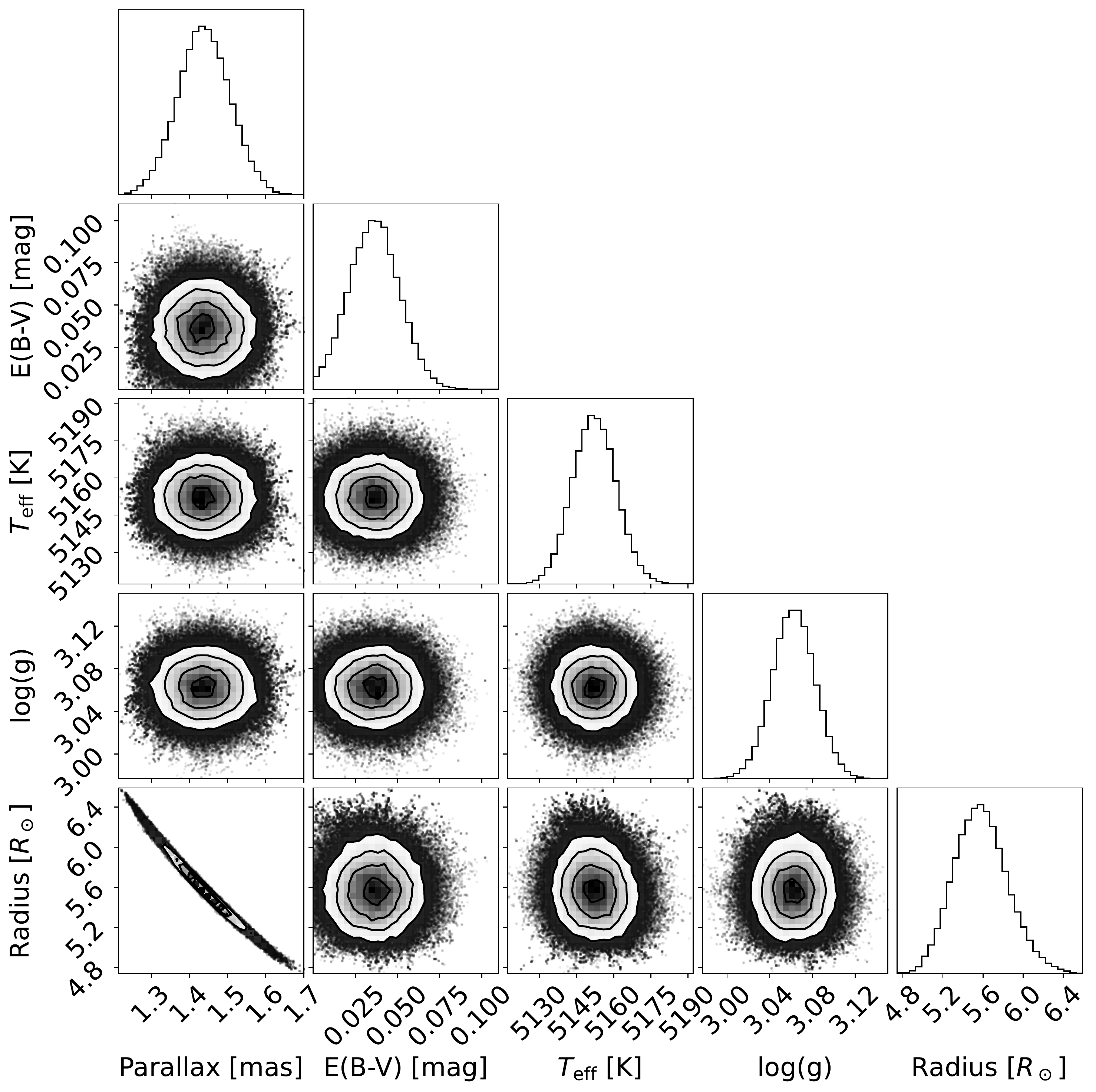}
    \caption{Posterior probability distributions for model parameters obtained through fitting the SEDs of 2MASS\,J1836$-$5110 (left), TYC\,6992$-$827$-$1 (centre) and TYC\,8394$-$1331$-$1 (right).}
  \label{fig:sed_corner}
  \end{center}
\end{figure*}

\section{WD+AFGK binaries from The White Dwarf Binary Pathways Survey}

\begin{landscape}
\begin{table}
 \centering
  \caption{Published WD+AFGK binaries with measured orbits discovered in The White Dwarf Binary Pathways Survey. References: (1) \citet{Parsons15}, (2) \citet{Hernandez22}, (3) \citet{Hernandez21}, (4) \citet{Hernandez22b}, (5) this paper. }
  \label{tab:wd+afgk}
  \tabcolsep=0.15cm
  \begin{tabular}{@{}lccccccccccccc@{}}
    \hline
    Name & P$_\mathrm{orb}$ [d] & T$_\mathrm{eff,WD}$ [K] & $\log{g}_\mathrm{WD}$ [dex] & M$_\mathrm{WD}$ [\MSUN] & T$_\mathrm{eff,2}$ [K] & $\log{g}_\mathrm{2}$ [dex] & M$_\mathrm{2}$ [\MSUN] & R$_\mathrm{2}$ [\RSUN] & $[\mathrm{M/H}]$ [dex] & $a$ [\RSUN] & $i$ [deg] & Distance [pc] & Reference \\
    \hline
TYC\,6760$-$497$-$1  & 0.49869(3) & $20250\pm750$ & $7.95\pm0.15$ & $0.60\pm0.08$ & $6400\pm100$ & $4.40\pm0.08$ & $1.24\pm0.02$ & $1.29\pm0.11$ & $-0.28\pm0.03$ & $3.24\pm0.04$ & $38\pm5$ & $295.2\pm2.6$ & (1) \\
TYC\,110$-$755$-$1   & 0.85805(1) & $16850\pm35$ & $8.39\pm0.01$ & $0.78\pm0.03$ & $5560\pm13$ & $4.24\pm0.05$ & $0.80\pm0.09$ & $1.114\pm0.006$ & $-0.14\pm0.14$ & $4.43\pm0.09$ & $20\pm2$ & $134.2\pm0.3$ & (2) \\
TYC\,4962$-$1205$-$1 & 1.2798(26) & - & - & $0.59-0.77$ & $5380\pm30$ & $4.13\pm0.03$ & $0.97\pm0.06$ & $1.404\pm0.012$ & $-0.42\pm0.07$ & $5.4-5.6$ & $37-47$ & $77.1\pm0.3$ & (3) \\
CPD$-$65 264         & 1.3704(1) & $24600\pm50$ & $8.38\pm0.01$ & $0.87\pm0.01$ & $5950\pm30$ & $4.39\pm0.02$ & $1.00\pm0.05$ & $1.06\pm0.01$ & $-0.14\pm0.05$ & $6.39\pm0.06$ & $64\pm1$ & $206.0\pm0.5$ & (4) \\
TYC\,1380$-$957$-$1  & 1.613(11) & - & - & $0.64-0.85$ & $5815\pm65$ & $4.41\pm0.05$ & $1.18\pm0.15$ & $1.122\pm0.014$ & $-0.03\pm0.07$ & $7.1-7.3$ & $49-71$ & $163.1\pm1.3$ & (3) \\
TYC\,3858$-$1215$-$1 & 1.6422(8) & - & - & $0.21-0.68$ & $4410\pm10$  & $4.55\pm0.04$ & $0.61\pm0.07$ & $0.679\pm0.004$ & $+0.15\pm0.37$ & $5.5-6.4$ & $25-87$ & $68.4\pm0.9$ & (2) \\
TYC\,4700$-$815$-$1  & 2.4667(87) & - & - & $0.38-0.44$ & $6040\pm31$ & $3.92\pm0.05$ & $1.45\pm0.17$ & $2.190\pm0.023$ & $-0.04\pm0.13$ & $9.4-9.5$ & $63-90$ & $165.8\pm1.2$ & (3) \\
TYC\,6992$-$827$-$1  & 41.45(1)  & $15750\pm50$  & $7.14\pm0.02$ & $0.28\pm0.01$ & $5250\pm50$ & $3.48\pm0.04$ & $1.31\pm0.14$ & $3.45\pm0.12$ & $-0.10\pm0.10$ & $58.6\pm1.8$ & $26\pm2$ & $500\pm20$  & (5) \\
TYC\,8394$-$1331$-$1 & 51.851(9) & $19400\pm100$ & $6.53\pm0.03$ & $0.24\pm0.01$ & $5150\pm20$ & $3.06\pm0.02$ & $1.31\pm0.12$ & $5.57\pm0.24$ & $+0.03\pm0.15$ & $67.7\pm0.3$ & $39\pm2$ & $680\pm30$ & (5) \\
2MASS\,J1836$-$5110  & 461.48(4) & $22250\pm250$ & $7.49\pm0.03$ & $0.40\pm0.01$ & $5050\pm50$ & $3.48\pm0.05$ & $1.38\pm0.16$ & $3.54\pm0.07$ & $-0.05\pm0.07$ & $304\pm9$ & $47\pm4$ & $900\pm20$ & (5)\\
    \hline
  \end{tabular}
\end{table}
\end{landscape}

\bsp
\label{lastpage}
\end{document}